\documentclass{article}

\usepackage[english]{babel}

\usepackage{indentfirst}
\usepackage{amsmath, amssymb, mathrsfs}

\usepackage{verbatim}

\usepackage[normalem]{ulem}
\usepackage{physics}

\usepackage{graphicx}
\graphicspath{{pictures/}}
\usepackage{subcaption}

\usepackage{array}
\usepackage{multirow}
\usepackage{hhline}
\usepackage{makecell}

\usepackage{enumitem}

\usepackage[
    backend=biber,
    natbib=true,
    style=numeric-comp,
    sorting=none]{biblatex}
\usepackage{csquotes}
\usepackage{textcase}
\usepackage{xcolor}

\usepackage[colorlinks,urlcolor=blue]{hyperref}

\usepackage{geometry}
\newcommand{\vpr}{\mkern1.5mu\raisebox{1.25ex}{\hbox{\rule{0.080em}{0.75ex}}}}
\newcommand{\vvpr}{\mkern1.5mu\raisebox{1.25ex}{%
  \hbox{\rule{0.080em}{0.75ex}\kern0.08em\rule{0.080em}{0.75ex}}}}

\begin{document}
\title{Spontaneous baryogenesis with large misalignment}

\author{Maxim Krasnov$^{1,2}$ \and Ufuk Aydemir$^3$ \and Maxim Khlopov$^4$ }
\date{%
    $^1$National Research Nuclear University MEPhI, Moscow 115409, Russia\\%
    $^2$Research Institute of Physics, Southern Federal University, Rostov-on-Don 344090, Russia\\
    $^3$Department of Physics, Middle East Technical University, Ankara 06800, T\"urkiye\\
    $^4$Virtual Institute of Astroparticle Physics, Paris 75018, France\\[2ex]%
    \today
}

\maketitle
\begin{abstract}
We investigate particle production by a pseudo-Nambu-Goldstone boson (pNGB) in
the spontaneous baryogenesis scenario for large misalignment angles. Since the
fermionic backreaction is intrinsically nonlocal in time, the large-angle problem is in
general difficult to treat directly. We argue that the adiabaticity conditions
are parametrically satisfied in the model, allowing the backreaction to be
described by a local Markovian approximation while retaining the nonlinear
dependence of the pNGB potential on the angular field. Through a numerical
study of arbitrary initial phases, we reproduce the cubic dependence of the
baryon asymmetry for small oscillations and demonstrate that this behavior
breaks down for large oscillations, especially for initial phases close to
$\pi$. Our calculations indicate that particle production saturates as the
initial phase approaches $\pi$ in Minkowski spacetime. The analysis is then extended to conformal Friedmann--Lemaître--Robertson--Walker (FLRW) spacetime, where the generated asymmetry shows a pronounced dependence on the damping rate of the pNGB motion. We further discuss the baryon-isocurvature bound on the ratio $H_\star/f$ and present sample parameter sets that satisfy this constraint at large misalignment. We also discuss the probability distribution of the baryon asymmetry.
\end{abstract}

\section{Introduction}
\label{introduction}

It is observationally evident that we live in a baryon-asymmetric universe even though there appears to be no \textit{a priori} reason for an asymmetry between particle and antiparticle production.  The baryon asymmetry is commonly expressed in terms of the current value of the baryon-to-entropy-density ratio
 $(\Delta n_{\mathrm{B}}/s)_0 \simeq 8.6\times 10^{-11}$ ~\cite{Planck:2018vyg}. Explaining this value naturally through a physical mechanism, instead of just as an initial condition at the beginning of the universe, has been a challenging task for decades~\cite{Barrow:2022gsu}. The idea of baryogenesis proposed by Sakharov and then Kuzmin \cite{Sakharov:1967dj, KUZMIN} related the generation of baryon excess in the initially baryon-symmetric universe to the nonequilibrium effects of CP violation in baryon nonconserving processes. The subsequent development of this idea resulted in various mechanisms linking the origin of the baryon asymmetry of the universe to physics beyond the Standard Model (SM).
 
A mechanism to generate baryon asymmetry spontaneously was proposed in Refs.~\cite{Cohen:1987vi,Cohen:1988kt} and was studied further in Refs.~\cite{Dolgov_1995,Dolgov_1997} (see Refs.~\cite{Barenboim_2019, Simone_2017, Simone_2016} for more recent studies and Refs.~\cite{Khlopov:2004rw,Dolgov:1991fr} for reviews). There is also similar gravitational baryogenesis \cite{PhysRevLett.93.201301} model, in which baryonic current is coupled to the derivative of the scalar curvature $\partial_\mu R$. However, this mechanism is unstable as shown in \cite{Arbuzova:2017zby}. Such coupling with $\partial_\mu R$ leads to explosive growth of the scalar curvature. In the spontaneous baryogenesis scenario, the asymmetry is produced through the relaxation of the (pseudo) Nambu-Goldstone boson (i.e. the phase $\theta=\phi/f$) of a spontaneously broken global $U(1)$ baryonic symmetry down to its potential, where $f/\sqrt{2}$ is the radial part of the vacuum expectation value of the complex scalar field responsible for the symmetry breaking. This field is a spectator at the inflation stage that coexists with the inflaton. The potential is tilted due to an explicit symmetry breaking, described by the term\footnote{This is similar to the QCD-axion scenario except in this case the potential is not produced from the QCD instantons effects.} $V(\theta)=\Lambda^4(1-\cos\theta)$, which also introduces mass to the originally massless Nambu-Goldstone boson.  The field $\theta$ is derivatively coupled to the non-conserved baryonic current, namely through the dimension-5 operator~$\mathcal{L}_{\mathrm{B}}=f^{-1}J_{\mathrm{B}}^{\mu}\partial_\mu \phi$, where $J_{\mathrm{B}}^{\mu}= \overline{Q}\gamma^\mu Q$ with $Q$ being a new heavy fermion that carries baryon number. The field $\theta$, through damped oscillations, is converted to baryons or antibaryons depending on the direction in which it is rotated toward the minimum of the slightly tilted Mexican hat potential. Depending on the initial angle $\theta_{in} $, the baryon asymmetry is generated.

In this paper, we investigate the effects of large initial misalignment angles
in the spontaneous baryogenesis scenario. The small-angle approximation
considered in the literature is useful and informative, but the distribution of
phase values at the end of inflation does not guarantee that the initial phase
is small. It is therefore important to understand the effects of large
misalignment angles. The region near $\theta_{\rm in}\simeq\pi$ is especially
interesting, since it corresponds to the local maximum of the potential.
Depending on whether the phase rolls down clockwise or counterclockwise, it
relaxes to a minimum at $\theta=0$ or $\theta=2\pi$, respectively. Thus,
$\theta_{\rm in}=\pi$ corresponds to the domain wall separating two degenerate
vacua. In our analysis, we start the motion close to this point,
$\theta_{\rm in}\simeq\pi$, in order to investigate its impact on the
generation of the baryon asymmetry.

A direct treatment of this large-angle regime is complicated by the fact that
the fermionic backreaction is intrinsically nonlocal. The evolution of the
coherent pNGB mode is then governed by a memory integral rather than by a
purely local equation of motion. In this work, we use a local Markovian
description of the dissipative part of this backreaction. This approximation is
justified when the pNGB background varies slowly compared with the microscopic
time scales entering the fermionic response. As we will show, the corresponding adiabaticity conditions are parametrically satisfied as a consequence of the scale hierarchy built into the model. The nonlocal backreaction can therefore be represented by a local dissipative term,
while the nonlinear dependence of the pNGB potential on the angular field is
kept explicitly. Such local reductions have been studied both at zero
temperature and in thermal backgrounds~\cite{Berera:2001gs,Moss:2006gt,Berera:2008ar,Buldgen:2019dus}. This effective description allows us to treat the post-inflationary evolution of the pNGB
field for arbitrary initial phases.

There is a possibility that cosmological inflation is driven by the
Nambu--Goldstone boson in question, a scenario referred to as natural
inflation~\cite{Freese:1990rb}. However, recent studies indicate that natural
inflation is strongly disfavored~\cite{Alam_2024,Montefalcone_2023,dos_Santos_2024};
it is in tension with Planck constraints on the tensor-to-scalar ratio $r$ and
the scalar spectral index $n_s$~\cite{Planck:2018jri}. In the model considered in this
paper, the Nambu--Goldstone boson responsible for baryogenesis is instead
taken to be a spectator field during inflation, and we remain agnostic about
the mechanism driving inflation. The pNGB is assumed to be present during
inflation, but its classical dynamics are frozen, while quantum fluctuations
populate different initial phase values.

The paper is organized as follows. In Section~\ref{model}, we review the
spontaneous baryogenesis model and its pNGB degree of freedom. In
Section~\ref{distribution}, we discuss the probability distribution of the
initial phase. In Section~\ref{stationary}, we study the evolution of
$\theta$ in Minkowski spacetime. In Section~\ref{FLRW}, we extend the analysis
to conformal FLRW spacetime. In Section~\ref{baryonasymmetry}, we calculate
the resulting baryon asymmetry and study its dependence on the initial phase.
In Section~\ref{sec:iso}, we derive the baryon-isocurvature bound associated
with inflationary fluctuations of the initial phase. Finally, we summarize our
results and discuss possible implications. We adopt the convention
$c=\hbar=k_B=1$ throughout this work, unless otherwise stated.

\section{The Model\label{model}}

We begin by laying out the basics of the spontaneous baryogenesis model, following the foundational works by A. Dolgov and his collaborators~\cite{Dolgov_1995, Dolgov_1997}. The core mechanism involves a complex scalar field $\Phi$ which, as we will see below, acquires a vacuum expectation value (VEV), triggering spontaneous symmetry breaking and giving rise to a Nambu–Goldstone boson. This Nambu–Goldstone mode then drives the generation of baryon number asymmetry. The model also contains two fermionic fields, $Q$ and $L$, with $Q$ carrying the SM baryon charge and $L$ taken to be neutral under the SM baryon and lepton charges. Contrary to what the notation may suggest, these fields are not SM–like quarks or leptons; they are singlets under the SM gauge group. 
The Lagrangian is given as
\begin{equation}
\label{UnbrokenL}
\mathcal{L}=\partial_\mu \Phi^* \partial^\mu \Phi - V(\Phi^*\Phi)+i\overline{Q}\gamma^\mu \partial_\mu Q+i\overline{L}\gamma^\mu \partial_\mu L - m_Q \overline{Q}Q -
    m_L \overline{L}L + g(\Phi \overline{Q} L+ \Phi^* \overline{L}Q).
\end{equation}
The Yukawa interaction term, $g(\Phi \overline{Q} L + \Phi^* \overline{L}Q)$, is of crucial importance as it will later facilitate the production of the $Q$ field, violating baryon number.
This Lagrangian possesses a classical global $U(1)$ symmetry associated with baryon number, under which the fields transform as:
\begin{equation}
\label{baryonsymmetry}
\Phi \rightarrow e^{i\alpha}\Phi, \quad Q \rightarrow e^{i\alpha}Q, \quad L \rightarrow L.
\end{equation}

The scalar potential $V(\Phi)$ is chosen to trigger spontaneous symmetry breaking (SSB) of this $U(1)$ at the energy scale $f$:
\begin{equation}
V(\Phi)=\lambda \left(\Phi^*\Phi - f^2/2\right)^2,
\end{equation}
which induces a nonzero vacuum expectation value,
$\langle\Phi\rangle=(f/\sqrt{2})e^{i\delta}$. The arbitrary constant phase
$\delta$ labels the degenerate vacuum manifold, and choosing a particular value
of it spontaneously breaks the $U(1)$ symmetry.  Expanding around the symmetry-breaking vacuum, the radial mode acquires a mass of order $m_r\sim \sqrt{\lambda}f$, while the angular degree of freedom remains massless as the Nambu-Goldstone boson. For processes below the symmetry-breaking scale, the heavy radial mode can be integrated out. Parameterizing the remaining low-energy field as $\Phi(x) = (f/\sqrt{2}) e^{i\theta(x)}$, where $\theta(x) \equiv \phi(x)/f$, and substituting into the original Lagrangian yields the effective theory below the SSB scale:
\begin{equation}\label{AftSymBroken2}
\mathcal{L}=\frac{f^2}{2}\partial_\mu \theta \partial^\mu \theta + i\overline{Q}\gamma^\mu \partial_\mu Q + i\overline{L}\gamma^\mu \partial_\mu L - m_Q \overline{Q}Q - m_L \overline{L}L +
    \frac{gf}{\sqrt{2}}\left(\overline{Q} L e^{i\theta} + \overline{L}Q e^{-i\theta}\right).
    \end{equation}
Notice that this Lagrangian remains invariant under the shifted $U(1)$ transformation given as
    \begin{equation}\label{transformations}
    Q\rightarrow e^{i\alpha}Q, \quad L \rightarrow L, \quad \theta \rightarrow \theta + \alpha.
    \end{equation}
In order to generate mass for the $\theta$ field and provide it with a potential to roll down, an explicit symmetry-breaking term can be introduced. To this end, the following potential is added to the Lagrangian above. 
\begin{equation}
\label{cosinepotential}
V(\theta) = \Lambda^4(1 - \cos\theta).
\end{equation}
This is the axion potential generated by instantons in QCD but is considered here as a generic low-energy effect parameterized by a scale $\Lambda \ll f$.
This potential tilts the initial Mexican hat potential, giving the pseudo-Nambu-Goldstone boson a mass $m_\theta \sim \Lambda^2/f$. 
One can express the Lagrangian, given in Eq.~(\ref{AftSymBroken2}), in an alternative form by applying the field redefinition $Q\rightarrow e^{-i\theta(x)}Q$. This rotation removes the phase from the Yukawa term and the term with derivative coupling arises:
\begin{equation}\label{AftSymBroken3}
\mathcal{L}=\frac{f^2}{2}\partial_\mu \theta \partial^\mu \theta + i\overline{Q}\gamma^\mu \partial_\mu Q + i\overline{L}\gamma^\mu \partial_\mu L - m_Q \overline{Q}Q - m_L \overline{L}L + \frac{gf}{\sqrt{2}}(\overline{Q} L + \overline{L}Q) + \partial_\mu\theta  \overline{Q}\gamma^\mu Q - V(\theta).
    \end{equation}  
    
The term $\partial_\mu\theta\,\overline{Q}\gamma^\mu Q$ is the hallmark of
spontaneous baryogenesis. It represents a derivative coupling between the
background $\theta$ field and the baryon current
$J_{\mathrm B}^{\mu}=\overline{Q}\gamma^\mu Q$. When the background is
time-dependent, $\dot\theta\neq0$, this coupling distinguishes the two
directions of motion in field space and biases baryon- and antibaryon-producing
processes. In this sense, $\dot\theta$ plays the role of the time-dependent
source that controls the generated baryon number density.\footnote{The quantity $\dot\theta$ is often described as an effective
chemical potential for baryon number, following the standard terminology of
spontaneous baryogenesis~\cite{Cohen:1987vi,Cohen:1988kt}. This
should not be understood as a basis-independent literal identification in all
formulations; in particular, its Hamiltonian interpretation can be
subtle~\cite{Arbuzova_2016,Arbuzova:2017zby}. This caveat does not affect the
results of Refs.~\cite{Dolgov_1995,Dolgov_1997}, where the baryon density is
obtained from the dynamical evolution of the coupled $\theta$--fermion system.
In the present work, we use this dynamical framework rather than relying on a
literal chemical-potential interpretation.} This bias is converted into a net baryon asymmetry during the
production of the heavy $Q$ fermions, mediated by the Yukawa interaction
$g(\overline{Q}L e^{i\theta}+\overline{L}Q e^{-i\theta})$. Thus, the
Lagrangian in Eq.~(\ref{AftSymBroken3}) provides a minimal framework for
spontaneous baryogenesis driven by a rolling pseudo-Nambu--Goldstone boson.

For the misalignment picture to apply, the $U(1)$ symmetry must be broken
before or during inflation, so that the radial expectation value is larger
than the inflationary fluctuations, parametrically $f\gtrsim H_\star$. If this
condition is not satisfied, fluctuations of $\Phi$ can exceed its expectation
value, the symmetry is effectively restored during inflation, and the phase is
not a well-defined light degree of freedom with a single value in each Hubble
patch.

Before closing this section, we make a few additional comments. A field redefinition similar to that applied to obtain Eq.~\eqref{AftSymBroken3}
can also be made on the $L$ field, namely $L\rightarrow e^{i\theta(x)}L$.\footnote{This issue was first brought to our attention by an anonymous referee.}
This redefinition shifts the derivative coupling to the $L$ current,
$\partial_\mu\theta\,\overline{L}\gamma^\mu L$, where
$\overline{L}\gamma^\mu L\equiv J_{\mathrm L}^{\mu}$.
It does not change any physical observable, since it is only a local change
of field variables. This can also be seen from the equations of motion for
the $Q$ and $L$ fields, which are given by~\cite{Dolgov_1995}
\begin{equation}
\label{currents}
\begin{gathered}
\partial_\mu\left(\bar{Q}\gamma^\mu Q\right)
\equiv \partial_\mu J_B^\mu
= -\frac{i g f}{\sqrt{2}}(\bar{Q}L-\bar{L}Q)
= \partial_\mu J_L^\mu ,
\\
\partial^\mu\partial_\mu\theta
+\frac{1}{f^2}\frac{dV(\theta)}{d\theta}
= -\frac{1}{f^2}\partial_\mu J_B^\mu
= \frac{ig}{\sqrt{2}f}(\bar{Q}L-\bar{L}Q)
= -\frac{1}{f^2}\partial_\mu J_L^\mu .
\end{gathered}
\end{equation}
As one can see, the nonconservation of the two currents is governed by the
same Yukawa interaction. The two descriptions are therefore related by a
field redefinition and give the same physical baryon asymmetry.

The equality of the two divergences does not, however, imply that the same
amount of SM lepton asymmetry is produced. This depends on whether the new
field $L$ carries SM lepton number. If the additional global $U(1)$
transformation
\[
\Phi \to e^{i\beta}\Phi, \qquad Q\to Q,\qquad L\to e^{-i\beta}L
\]
is identified with SM lepton number, then the total SM $B-L$ produced by
the model vanishes. This is important because electroweak sphalerons, which
are active in thermal equilibrium for
$T_{\mathrm{ew}}\sim 10^{2}\,\mathrm{GeV}\lesssim T\lesssim 10^{12}\,\mathrm{GeV}$,
violate $B+L$ while conserving $B-L$. If the total $B-L$ vanishes,
these processes drive $B+L\to 0$, thereby washing out any preexisting
baryon asymmetry~\cite{Harvey:1990qw,Buchmuller:2000as}. In this case,
baryon asymmetry must be generated after sphaleron freeze-out, namely
$T_{\mathrm{osc}}\lesssim T_{\mathrm{ew}}$.

However, if the $U(1)$ symmetry above is not identified with the SM lepton number, no SM lepton asymmetry is generated. In that
case, the oscillating $\theta$ field produces a nonzero SM $B-L$, which
is conserved by sphalerons. Therefore, the generation of baryon asymmetry is
not subject to sphaleron washout and can occur above the electroweak scale,
$T_{\mathrm{osc}}>T_{\mathrm{ew}}$, as also noted in
Refs.~\cite{Dolgov_1995,Dolgov_1997}. In this paper, we adopt this
approach.\footnote{Such a sphaleron-washout
concern was raised in Ref.~\cite{Cline-published}. The charge assignment
specified here clarifies why this issue does not arise.}

\section{Distribution of the initial phase\label{distribution}}

The initial value of the phase field $\theta_{in}$ at the beginning of its oscillations is not a fixed parameter and determined by quantum fluctuations during cosmological inflation. We consider the probability distribution $f(\phi, t)$ for a light scalar field $\phi$ (where $\theta = \phi/f$) during inflation. This distribution could be obtained by solving the Fokker-Planck equation \cite{LINDE1982335, Starobinsky:1986fx, Vennin_2015}, which, for a massless field (where $m \ll H_\star$), yields a Gaussian distribution. Starting from an initial value $\phi_u$ at the start of the inflation, the probability density to find the field at a value $\phi$ after a time $t$ is given as~\cite{Belotsky:2018wph}
\begin{equation}\label{probDensity}
    f(\phi,t) = \frac{1}{\sqrt{2\pi}\,\sigma(t)} \exp\left(-\frac{(\phi - \phi_u)^2}{2\sigma^2(t)}\right),
\end{equation}
where $\sigma(t) = \frac{H_\star}{2\pi} \sqrt{H_\star t}$. This represents the random walk of the field value due to quantum fluctuations superimposed on the classical slow roll.
The baryon asymmetry generated in the spontaneous baryogenesis mechanism is highly sensitive to the initial phase $\theta_{in}$ at the end of inflation. Converting the distribution for $\phi$ into one for the phase $\theta_{in} = \phi_i / f$, and assuming inflation lasts for $N \approx 60$ e-folds ($t \approx 60 H_\star^{-1}$), we obtain the probability distribution for the initial misalignment angle after inflation:
\begin{equation}\label{distr}
    f(\theta_{in}) = \frac{1}{\sqrt{2\pi}\,\sigma'} \exp\left(-\frac{(\theta_{in} - \theta_u)^2}{2\sigma'^2}\right),
\end{equation}
where $\sigma' = \frac{H_\star}{2\pi f} \sqrt{60}$.
A key feature of cosmological inflation is that causally disconnected regions evolve independently. The entire observable universe today originates from approximately $e^{3N} \approx e^{180}$ such independent Hubble patches at the end of inflation. Within each patch, the value of $\theta_{in}$ is nearly constant, but varies randomly from patch to patch according to the distribution \eqref{distr}. This makes spontaneous baryogenesis an \textit{inhomogeneous} process on super-Hubble scales at this epoch; different regions will produce different baryon asymmetries.
The probability that a given Hubble patch has a misalignment angle shifted by more than $\pi$ from its initial value $\theta_u$ is given as~\cite{Belotsky:2018wph}
\begin{equation}
    P(|\theta_{in} - \theta_u| > \pi) = 1 - \text{erf}\left(\frac{\pi}{\sqrt{2}\sigma'}\right).
\end{equation}
Assuming the symmetry breaking scale $f$ is comparable to the Hubble scale during inflation ($f \approx H_\star$)\footnote{We take $f\approx H_\star$ here only for the sake of this estimate,
since it gives the largest accumulated dispersion while remaining at the edge
of the misalignment regime introduced above, $f\gtrsim H_\star$. The
isocurvature constraint discussed in Sec.~\ref{sec:iso} imposes the
stronger hierarchy $H_\star\ll f$. In that regime, inflationary fluctuations
of the phase are small, so $\theta_{\rm in}$ is effectively homogeneous across
the observable universe. Its homogeneous value is nevertheless an arbitrary
angle on the vacuum manifold; hence values near $\pi$ are as natural as any
other initial misalignment angle.}, we find $\sigma' \approx \sqrt{60}/(2\pi) \approx 1.23$, and thus:
\begin{equation}
    P(|\theta_{in} - \theta_u| > \pi) \approx 1 - \text{erf}(\pi) \approx 10^{-5}.
\end{equation}
Although this probability for a single patch is small, the total number of patches is very large. The expected number of patches within our observable universe that have experienced such a large fluctuation is:
\begin{equation}
    n_{\text{regions}} = e^{180} \times P(|\theta_{in} - \theta_u| > \pi) \approx 10^{78} \times 10^{-5} \gg 1.
\end{equation}
Therefore, it is virtually certain that regions with $\theta_{in} \sim \pi$ exist within our modern horizon. This requires thorough investigation of the baryogenesis mechanism for these large initial misalignment angles, which is the primary focus of this work.
\\

\section{Evolution of $\theta$ in Minkowski spacetime\label{stationary}}
\subsection{Local reduction of the nonlocal backreaction\label{sec:local}}
In this section, we consider the equation of motion in Minkowski spacetime and
work in the simplifying limit in which the fermion masses vanish. For an
arbitrary initial phase, the relevant semiclassical equation of motion is
given by~\cite{Dolgov_1995}:
\begin{eqnarray}
\label{nonlocal_eom}
    \Ddot{\theta}+\cfrac{\Lambda^4}{f^2}\sin{\theta} &=& -\cfrac{4g^2}{\pi^2} \int_0^\infty \omega^2 d\omega   \int_{-\infty}^0 dt'\;\sin{(2\omega t')}\;\sin{[\theta(t+t')-\theta(t)]}\nonumber\\
    &\equiv& \mathcal{I}(t) .
\end{eqnarray}
This equation is obtained by treating the scalar field $\theta$ classically
and the fermion fields $Q$ and $L$ quantum mechanically. 

Equation~(\ref{nonlocal_eom}) should be understood as a formal expression for the fermionic backreaction within the low-energy pNGB description. In writing it, we have not specified an explicit ultraviolet regulator for the frequency integral. As discussed in Sec.~\ref{model}, the radial mode associated with the symmetry-breaking field is heavy and has been integrated out. The resulting theory for the angular mode $\theta$ is therefore an effective pNGB theory, valid only below the scale at which this description breaks down. We denote this microscopic scale by $\omega_{\rm cut}$, which is expected to be of order the symmetry-breaking scale, $\omega_{\rm cut}\sim f$.\footnote{Ref.~\cite{Cline-published} characterizes the cutoff scale as
introduced without justification. In the present effective description, this scale is not an additional assumption tied specifically to the memory integral,
but rather the ultraviolet breakdown scale of the low-energy pNGB theory. After the
radial mode is integrated out, the remaining angular degree of freedom is
described by an EFT whose interactions are suppressed by the symmetry-breaking
scale. This is reflected, for example, in the derivative coupling
$(\partial_\mu\phi/f)\,\overline{Q}\gamma^\mu Q$ in
Eq.~(\ref{AftSymBroken3}), where $\phi$ denotes the canonically normalized
pNGB field.} The same scale sets the microscopic memory time of the effective description,
$t_{\rm mem}\sim \omega_{\rm cut}^{-1}$. Thus, the memory kernels should be
interpreted as EFT kernels whose short-time structure is coarse-grained over
time separations of order $t_{\rm mem}$, rather than as kernels resolving
arbitrarily short-time dynamics.

In the small-angle approximation, the nonlocal backreaction, or memory, term on
the right-hand side, denoted by $\mathcal I(t)$, generates a local
dissipative contribution proportional to $\dot\theta$~\cite{Dolgov_1995,Dolgov_1997}. For large initial angles, the
nonlocal term is much harder to treat directly. However, as we argue below, the
relevant adiabaticity conditions are parametrically satisfied in the model.
This allows the leading dissipative part of the backreaction to be represented
by the same type of local term. The resulting effective equation remains
tractable for arbitrary initial angles, while retaining the full nonlinear term
$\sin\theta$ on the left-hand side.\footnote{In an earlier version of this manuscript, locality in time was implemented through a delta-function approximation to the memory kernel as an intermediate step in the discussion of the local dissipative contribution. As
noted in Ref.~\cite{Cline-published}, that approximation is not a valid
delta-function identity, and we do not use it here. Instead, we implement the
Markovian limit by expanding the phase difference within the microscopic
memory time. As shown below, the leading term in this adiabatic expansion
yields a local dissipative contribution linear in $\dot\theta$, while the
subleading higher-derivative terms are suppressed by powers of the adiabatic
parameter. The resulting coefficient is treated as a transport parameter of
the local effective description.}
The local treatment of the dissipative backreaction can be understood as a
Markovian effective description of the nonlocal memory term. In this
description, the memory dependence is replaced by local terms in the effective
equation of motion. Its validity is controlled by adiabaticity: the memory time of the fermionic response must be short compared with the time scale on which the background evolves. In this regime, the dissipative part of
 the response admits a local effective description, with coefficients encoding the underlying microscopic physics~\cite{Berera:2001gs,Moss:2006gt,Berera:2008ar,Buldgen:2019dus}.

The form of the local terms is constrained by the structure of the
nonlocal backreaction. In Eq.~(\ref{nonlocal_eom}), the memory term depends on the
background only through the difference $\Delta\theta=\theta(t+t')-\theta(t)$, and not on the absolute value of $\theta$ itself. This structure reflects the shift symmetry of the fermionic
coupling. Therefore, in the adiabatic regime relevant for the Markovian
reduction, the backreaction is expected to generate derivative terms rather
than local potential-like terms depending only on $\theta$ itself.\footnote{This statement refers to the fermionic backreaction at the order
considered here. The pNGB shift symmetry is already explicitly broken by the
potential, and higher-order effects involving this explicit breaking can
generate potential-like corrections. These are distinct from the leading
dissipative part of the memory response. Moreover, a term proportional to
$\theta$ in earlier small-angle treatments~\cite{Dolgov_1995,Dolgov_1997} can arise by rewriting derivative corrections using the zeroth-order equation of motion, e.g. $\ddot\theta\to -m_\theta^2\theta$. Such a term should be viewed as a mass-renormalization effect within the small-angle treatment, rather than as a new potential generated by the leading
backreaction.} Higher time-derivative terms, as well as higher powers of $\dot\theta$, are
suppressed in the adiabatic expansion, since
each additional derivative brings a small ratio of the background time scale to
the microscopic memory time scale. These observations support parametrizing the leading local dissipative contribution to $\mathcal I(t)$ by a term linear in $\dot\theta$. We now show
how this form arises from the memory integral.

In Eq.~(\ref{nonlocal_eom}), the memory integral contains two
distinct time scales: the microscopic oscillatory scale carried by the kernel
factor $\sin(2\omega t')$, and the macroscopic time scale over which the
background $\theta(t)$ varies. A frequency component $\omega$ in the kernel probes
time separations $|t'|\sim\omega^{-1}$. Therefore, the highest frequencies retained in the
low-energy description set the shortest microscopic time scale resolved by the
memory kernel. This scale sets the microscopic memory time entering the
Markovian reduction,
\begin{equation}
t_{\rm mem}\sim\omega_{\rm cut}^{-1}.
\end{equation}
The validity of the adiabatic expansion requires a separation between this
memory time and the evolution time scale of the
background~\cite{Buldgen:2019dus},
\begin{equation}
\label{adiabatic-condition}
t_{\rm mem}\ll t_\theta,
\end{equation}
where $t_\theta\sim |\dot\theta|^{-1}$ denotes the characteristic time scale
over which the dimensionless field $\theta$ changes by an order-one amount.

Consequently, within the memory window relevant for the local Markovian
reduction, $|t'|\lesssim t_{\rm mem}$, the background varies only weakly, as
guaranteed by the adiabatic condition in Eq.~(\ref{adiabatic-condition}).
The phase difference can therefore be expanded as
\begin{equation}
\label{derivative-expansion}
\theta(t+t')-\theta(t)
=
\dot\theta(t)t'
+\frac12\ddot\theta(t)t'^2+\cdots .
\end{equation}
The leading term is small by the adiabaticity condition, while the second term
is suppressed relative to it by a further power of the same small parameter,
$t_{\rm mem}/t_\theta\sim|\dot\theta|t_{\rm mem}$. Explicitly,\footnote{This
estimate uses $|\ddot\theta/\dot\theta|\sim t_\theta^{-1}$. More precisely,
$|\dot\theta/\ddot\theta|$ defines the time scale $t_{\dot\theta}$ over which
$\dot\theta$ undergoes an order-one fractional change. For a slowly varying
oscillatory background, $t_{\dot\theta}$ is parametrically of the same order
as $t_\theta$; for example, if $\theta\sim\cos(\omega t)$, both scale as
$\omega^{-1}$.}
\begin{equation}
\left|
\frac{\ddot\theta\, t'^2}{\dot\theta\, t'}
\right|
\sim
\frac{|t'|}{t_\theta}
\lesssim
\frac{t_{\rm mem}}{t_\theta}\ll1 .
\end{equation}
and higher-order terms are suppressed by further powers of this adiabatic
parameter.

Therefore, at leading adiabatic order, the sine of the phase difference
appearing in Eq.~(\ref{nonlocal_eom}) may be expanded to first order. With a
frequency cutoff and restricting the Markovian contribution to the memory
window $-t_{\rm mem}<t'<0$,\footnote{The endpoint $t'=0$ requires no separate
regularization once the frequency integral is cut off at $\omega_{\rm cut}$,
since the kernel is regular there and vanishes linearly as $t'\to0$. The fact
that the microscopic short-time dynamics are not resolved in the EFT refers to
the coarse-grained description, i.e. the description of the memory response
within the effective pNGB theory for the slowly varying field, rather than to
removing the endpoint from the memory integral.} the right-hand side of
Eq.~(\ref{nonlocal_eom}) becomes
\begin{equation}
{\cal I}(t)
\supset
-\frac{4g^2}{\pi^2}\dot\theta(t)
\int_0^{\omega_{\rm cut}}d\omega\,\omega^2
\int_{-t_{\rm mem}}^{0}dt'\,
t'\sin(2\omega t') .
\end{equation}
The inner integral is
\begin{equation}
\int_{-t_{\rm mem}}^{0}dt'\,
t'\sin(2\omega t')
=
-\frac{t_{\rm mem}\cos(2\omega t_{\rm mem})}{2\omega}
+
\frac{\sin(2\omega t_{\rm mem})}{4\omega^2}.
\end{equation}
Therefore,
\begin{equation}
\label{integral-result}
{\cal I}(t)
\supset
-\frac{4g^2}{\pi^2}
\left[
-\frac{\omega_{\rm cut}}{4}
\sin(2\omega_{\rm cut}t_{\rm mem})
+
\frac{1-\cos(2\omega_{\rm cut}t_{\rm mem})}
{4t_{\rm mem}}
\right]\dot\theta(t).
\end{equation}
In the dissipative convention
\begin{equation}
\label{dissipative-term}
{\cal I}(t)=-\Gamma_{\rm eff}\dot\theta(t)+\cdots ,
\end{equation}
this prescription would correspond to a cutoff-dependent value of
$\Gamma_{\rm eff}$. Here, as discussed above, $\omega_{\rm cut}\sim f$
denotes the physical cutoff scale of the low-energy pNGB effective theory
rather than an arbitrary regulator. For
$t_{\rm mem}\sim\omega_{\rm cut}^{-1}$, the magnitude of
$\Gamma_{\rm eff}$ is generically of order $g^2\omega_{\rm cut}$, but its sign and magnitude depend on the precise cutoff prescription and on the definition of the coarse-graining window. In particular, a negative value obtained from this prescription should not be interpreted as physical anti-damping, since energy flows from the coherent mode into the produced fermion pairs and not conversely. Rather, it reflects the limitations of the sharp-cutoff estimate, where oscillatory terms can strongly affect the extracted numerical coefficient. Likewise, for special
choices of $\omega_{\rm cut}t_{\rm mem}$, the same prescription can even give
a vanishing coefficient.\footnote{For discussions of related issues, including prescription dependence in the Markovian limit, see
Ref.~\cite{Greiner:1996dx}.}  Exact cancellation of the linear term would require a tuned relation between $\omega_{\rm cut}$ and $t_{\rm mem}$ and is therefore not generic. Even in such
a tuned case, dissipation would not be absent; the leading dissipative
response would instead arise from higher odd powers of $\dot\theta$. We do not
pursue this tuned possibility here. The physical origin of the leading local dissipative term and the sign of the corresponding transport coefficient are discussed from an energy-balance
perspective in Appendix~\ref{app:energy_balance}. 

In terms of the dimensionless time variable $\bar{t}=\Lambda^2 t/f$, using
the estimate obtained in the previous paragraph, we obtain
\begin{equation}
\Gamma=\Gamma_{\rm eff}\frac{f}{\Lambda^2}
=
\mathcal{N}\,g^2\frac{f^2}{\Lambda^2},
\end{equation}
where $\mathcal{N}$ denotes the dimensionless factor associated with the
cutoff-scale implementation of the memory integral.  In the sharp-cutoff estimate above, this factor depends on the combination
$\omega_{\rm cut}t_{\rm mem}$ through oscillatory functions. The Markovian
argument fixes only the order of magnitude $t_{\rm mem}\sim \omega_{\rm cut}^{-1}$, but not the precise numerical relation between $t_{\rm mem}$ and $\omega_{\rm cut}^{-1}$. Since the remaining factors entering the local coefficient are also not fixed by the low-energy pNGB description, we treat $\Gamma$ as a free parameter. We then
investigate the large-angle dynamics as a function of this parameter in both
weakly and strongly damped regimes, $\Gamma<1$ and $\Gamma\gtrsim1$.

The phenomenological treatment of $\Gamma$ should not be understood as an
intrinsic indeterminacy of this coefficient in the theory. Rather, $\Gamma$,
or its dimensionful counterpart $\Gamma_{\rm eff}$, is an effective transport
coefficient of the local Markovian description. In the linearized small-angle
regime, Refs.~\cite{Dolgov_1995,Dolgov_1997} showed that the divergences in
the nonlocal equation can be absorbed into the pNGB mass renormalization, and
obtained a definite local damping term. Away from the small-angle
approximation, the nonlocal memory integral retains its full oscillatory and
nonlinear dependence on the phase, and an analogous extraction of a single
local transport coefficient requires a separate first-principles treatment of
the renormalized nonlocal dynamics. We do not attempt that calculation here.
Instead, we use the adiabatic expansion of the nonlocal memory term to identify
the leading local dissipative structure and treat the damping coefficient
phenomenologically, studying how the generated asymmetry depends on it once
the small-angle approximation is dropped. As shown below, the adiabaticity conditions required for this Markovian treatment are parametrically satisfied as a consequence of the hierarchy $\Lambda\ll f$ built into the model.

The corresponding leading local equation, obtained by retaining the first
term in the adiabatic expansion of the memory integral, is then
\begin{equation}
\label{Minkowskieom}
\theta\vvpr+\Gamma\theta\vpr+\sin\theta=0 \;,
\end{equation}
where, as introduced above, the upright prime denotes differentiation with
respect to the dimensionless time variable $\bar{t}=\Lambda^2 t/f$, and
$\Gamma=\Gamma_{\rm eff}f/\Lambda^2$ denotes the dimensionless damping
parameter.

\subsection{Adiabatic conditions for the Markovian description} 
We now examine the conditions under which the Markovian treatment introduced
above can be applied. As emphasized earlier, the background must vary slowly
over the memory time of the kernel. In the present setup, the cutoff-dependent
kernel oscillates on the microscopic time scale $\omega_{\rm cut}^{-1}$, and
time separations much longer than this scale do not contribute coherently to
the convolution with a slowly varying background. The relevant adiabatic
conditions are therefore governed by the variation of the background over the
memory time $t_{\rm mem}\sim \omega_{\rm cut}^{-1}$.

A quantitative formulation follows from the separation between two
characteristic time scales. The kernel resolves microscopic intervals of order
$\omega_{\rm cut}^{-1}$, whereas the background phase changes appreciably on
the instantaneous time scale $t_\theta\sim |\dot\theta|^{-1}$. A local
Markovian description is justified provided the background changes negligibly
over the microscopic memory time, which requires
$\omega_{\rm cut}^{-1}\ll |\dot\theta|^{-1}$. The resulting adiabatic condition is\footnote{In an FLRW background, the same
reasoning also requires the scale factor to vary slowly over the memory time,
giving $H/\omega_{\rm cut}\ll1$, where $H$ is the Hubble parameter.}
\begin{equation}
\label{conditionMinkowski}
\frac{|\dot\theta|}{\omega_{\rm cut}} \ll 1 .
\end{equation}
A conservative upper bound on the field velocity can be obtained by neglecting
friction and using conservation of the corresponding frictionless energy,
\begin{equation}
\frac12 f^2\dot\theta^2
+
\Lambda^4(1-\cos\theta)
=
\Lambda^4(1-\cos\theta_{\rm in}) .
\end{equation}
Here, $\theta_{\rm in}$ denotes the initial value of the coherent field, taken
with $\dot\theta_{\rm in}=0$; in the large-angle regime of interest it may lie
close to the hilltop, $\theta=\pi$. Therefore,
\begin{equation}
|\dot\theta|_{\rm max}
=
\frac{\Lambda^2}{f}
\sqrt{2(1-\cos\theta_{\rm in})}
\leq
\frac{2\Lambda^2}{f}
=
2m_\theta .
\end{equation}
Taking $\omega_{\rm cut}\sim f$, we find
\begin{equation}
\frac{|\dot\theta|_{\rm max}}{\omega_{\rm cut}}
\lesssim
2\frac{\Lambda^2}{f^2}\ll1 .
\end{equation}
Since the hierarchy $\Lambda\ll f$ is already built into the model
construction, the adiabatic condition in Eq.~(\ref{conditionMinkowski}) is parametrically satisfied: the field changes only by a small amount over the microscopic memory time. Thus, the leading dissipative Markovian reduction of the nonlocal backreaction is self-consistent, and the local effective equation of motion in Eq.~(\ref{Minkowskieom}) is applicable.

\subsection{Numerical results}
\begin{figure}[ht]
    \centering
    \begin{subfigure}{0.5\textwidth}
        \centering
        \includegraphics[width=1\linewidth]{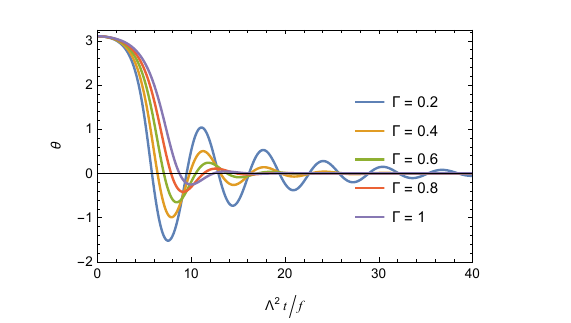}
        \caption{Numerical solutions for sample values of $\Gamma\leqslant1$.}
        \label{fig:a}
    \end{subfigure}%
    \begin{subfigure}{0.5\textwidth}
        \centering
        \includegraphics[width=1\linewidth]{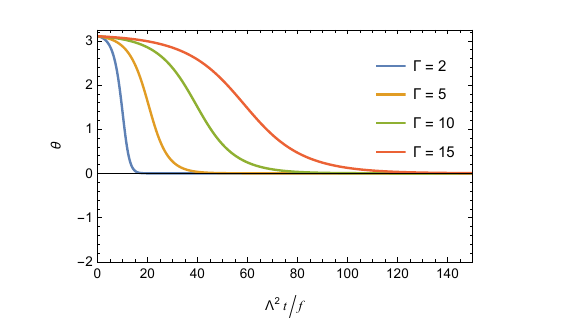}
        \caption{Numerical solutions for sample values of $\Gamma> 1$.}
        \label{fig:b}
    \end{subfigure}
    \caption{Numerical solutions of Eq.~\eqref{Minkowskieom} with initial conditions $\theta_{\text{in}}=3.1$ and $\dot{\theta}_{\text{in}} = 0$ for different values of $\Gamma$ on the Minkowski background.}
    \label{NumSol0}
\end{figure}

We present numerical solutions of Eq.~\eqref{Minkowskieom} in
Fig.~\ref{NumSol0} for different values of $\Gamma$, taking an initial phase
close to $\pi$. The exactly hilltop initial condition,
$\theta_{\rm in}=\pi$ with $\dot\theta_{\rm in}=0$, corresponds to an
unstable equilibrium point. With exact classical initial data it gives the
static solution $\theta(t)=\pi$, so the rolling solution is obtained by taking
$\theta_{\rm in}$ slightly away from $\pi$.

We allow both $\Gamma<1$ and $\Gamma\geq1$ in the numerical analysis. This
range is meant to explore both weakly and strongly damped local evolution, both
of which are compatible with the adiabaticity conditions derived above. The
adiabaticity conditions constrain the validity of the local Markovian
reduction, namely whether the background varies slowly over the microscopic
memory time of the kernel. By contrast, the size of $\Gamma$ controls the
character of the resulting local motion.

The results are displayed in two subfigures in order to make the plots more accessible. In contrast to the case of small oscillations, we see that the higher the value of $\Gamma$, the longer the motion takes to the bottom of the potential. This behavior is caused by large initial phase~--- potential term in the equation of motion behaves differently compared to small oscillations. 

\section{Evolution of $\theta$ in conformal FLRW spacetime}\label{FLRW}

\subsection{Equation of motion with nonlocal backreaction}
We now consider the spontaneous baryogenesis scenario in conformal FLRW background. Let us consider action in curved space-time:
\begin{multline}
    S=\int d^4x\sqrt{-g}\bigg[\frac{1}{2}f^2g^{\mu\nu}\partial_\mu\theta \partial_\nu\theta + i\overline{Q}\nabla_\mu \Gamma^\mu Q + \\ +i\overline{L}\nabla_\mu \Gamma^\mu L +g^{\mu\nu}\partial_\mu\theta \overline{Q}\Gamma_\nu Q -m_Q \overline{Q}Q -m_L\overline{L}L+\\+\frac{gf}{\sqrt{2}}(\overline{Q}Le^{i\theta}+\overline{L}Qe^{-i\theta})-V(\theta) \bigg].
\end{multline}

In conformal FLRW we have $ds^2=a^2(d\tau^2-d\mathbf{x}^2)$, thus we can obtain an effective Lagrangian:
\begin{multline}
    \mathcal{L} = \cfrac{1}{2}f^2a^2\partial_\mu\theta\partial^\mu\theta +i\overline{Q}\partial_\mu\gamma^\mu Q+i\overline{L}\partial_\mu\gamma^\mu L + \partial_\mu\theta \overline{Q}\gamma^\mu Q +\\+gfa(\overline{Q}Le^{i\theta}+\overline{L}Qe^{-i\theta})-a^4U(\theta) - am_Q \overline{Q}Q  - am_L \overline{L}L,
\end{multline}
where $a$ is the scale factor and fermions are redefined as $\psi \rightarrow \psi/a^{3/2}$ for the simplicity of the calculations.

To simplify our consideration, we will set fermion masses to be equal to zero. Thus, the Lagrangian we are considering is:
\begin{equation}
     \mathcal{L} = \cfrac{1}{2}f^2a^2\partial_\mu\theta\partial^\mu\theta +i\overline{Q}\partial_\mu\gamma^\mu Q+i\overline{L}\partial_\mu\gamma^\mu L + \partial_\mu\theta \overline{Q}\gamma^\mu Q + gfa(\overline{Q}Le^{i\theta}+\overline{L}Q e^{-i\theta})-a^4U(\theta) .
\end{equation}
In that case, the derivation is the same as in the Minkowski metric. The equation of motion is as follows \cite{Dolgov_1995}:
\begin{equation} 
\label{nonlocaleom}
\partial_\mu(a^2\partial^\mu\theta)+a^4\cfrac{\Lambda^4}{f^2}\sin{\theta}=-\cfrac{4g^2}{\pi^2}a(\tau)\int_0^\infty \omega^2 d\omega  \int_{-\infty}^0 a(\tau+\tau')\sin{(2\omega \tau')}\sin{[\theta(\tau+\tau')-\theta(\tau)]} d\tau',
\end{equation}
where $\tau$ is conformal time. As in the Minkowski case, the effective pNGB theory is valid only below the
symmetry-breaking scale $f$. Therefore, the UV cutoff of the effective theory
can be taken to be $\omega_{\rm cut}\sim f$, even though this cutoff is not
displayed explicitly in the formal expression above. The memory kernels should
therefore be understood as containing short-time structure characterized by
the microscopic scale
\begin{equation}
t_{\rm mem}\sim\omega_{\rm cut}^{-1},
\end{equation}
which sets the time resolution relevant for the local Markovian expansion.

\subsection{Adiabatic conditions for the Markovian reduction}
As discussed previously, the Markovian treatment requires the adiabatic regime where the background varies slightly over the characteristic time scale entering the memory kernel. For the Minkowski case, the condition is given in
Eq.~(\ref{conditionMinkowski}). In an FLRW spacetime, one must also require
the expansion of the universe to be slow on the same microscopic scale.
Therefore, in an FLRW spacetime, the adiabaticity conditions are
\begin{equation}
\label{conditionsFRLW}
\frac{|\partial_\tau\theta|}{a\,\omega_{\rm cut}}\ll 1,
\qquad
\frac{\mathcal H}{a\,\omega_{\rm cut}}\ll 1,
\end{equation}
where $\mathcal H\equiv a'/a=aH$ is the conformal Hubble parameter. The first
condition states that the pNGB background changes slowly on the physical
microscopic time scale $\omega_{\rm cut}^{-1}\sim f^{-1}$. Using
$\eta\equiv \Lambda^2\tau/f$, it can be written as
\begin{equation}
\frac{|\partial_\tau\theta|}{a\,\omega_{\rm cut}}
=
\frac{\Lambda^2}{f}
\frac{|\partial_\eta\theta|}{a\,\omega_{\rm cut}}
\sim
\frac{|\partial_\eta\theta|}{a}
\frac{\Lambda^2}{f^2}.
\end{equation}
For radiation domination, we choose the normalization $a(\eta)=\eta$. Evaluated
around the onset of the motion, $\eta=\eta_{\rm osc}$, this gives
\begin{equation}
\frac{|\partial_\tau\theta|}{a\,\omega_{\rm cut}}
\sim
\frac{|\partial_\eta\theta|}{\eta_{\rm osc}}
\frac{\Lambda^2}{f^2}.
\end{equation}
Since parametrically $\eta_{\rm osc}\sim1$, and since the numerical solutions
below show that $|\partial_\eta\theta|$ is at most of order unity around
$\eta_{\rm osc}$, this condition is controlled by the condition
$\Lambda^2/f^2\ll1$, as in the Minkowski case.

Similarly, the second condition in Eq.~(\ref{conditionsFRLW}), which concerns
the slow expansion of the universe over the microscopic memory time, gives
\begin{equation}
\frac{\mathcal H}{a\,\omega_{\rm cut}}
\sim
\frac{1}{\eta_{\rm osc}^2}
\frac{\Lambda^2}{f^2}
\ll 1 ,
\end{equation}
where we have again used $a(\eta)=\eta$ and $\omega_{\rm cut}\sim f$. Thus, this condition is also controlled by the small ratio
$\Lambda^2/f^2$. 

As a result, since the hierarchy $\Lambda/f\ll1$ is already built into the
model, the adiabaticity conditions are parametrically satisfied, allowing the nonlocal backreaction to be described by a local Markovian approximation.

\subsection{Effective local equation of motion}
Since the adiabaticity conditions are satisfied, the leading local
dissipative contribution from the backreaction in Eq.~(\ref{nonlocaleom}) can
be parametrized by a term proportional to $\Gamma_{\rm eff}\dot\theta$. Before
performing this local reduction, however, it is useful to isolate the slowly
varying scale-factor dependence in the memory integral. The nontrivial part of
the Markovian reduction is controlled by the rapidly oscillating kernel, while
the scale factor varies only on the Hubble time.

Because the kernel contains rapidly oscillating microscopic phases, its
coherent contribution to the local backreaction is controlled, after coarse
graining, by physical time separations of order $\omega_{\rm cut}^{-1}$. In
conformal time, this corresponds to
$|\tau'|\sim (a\;\omega_{\rm cut})^{-1}$. Over this memory time, the condition
$\mathcal H/(a\;\omega_{\rm cut})\ll1$ leads to
\begin{equation}
a(\tau+\tau')
=
a(\tau)\left[1+O\!\left(\frac{\mathcal H}{a(\tau)\;\omega_{\rm cut}}\right)\right].
\end{equation}
Thus, to leading adiabatic order, the factor $a(\tau+\tau')$ may be replaced
by $a(\tau)$ inside the memory integral.

The right-hand side of
Eq.~(\ref{nonlocaleom}) then reduces to the local dissipative term
$-a^2(\tau)\Gamma_{\rm eff}\dot\theta$ through the same procedure followed in the Minkowski case (through Eqs.~(\ref{derivative-expansion}) and (\ref{dissipative-term})).
Therefore, under adiabatic conditions, Eq.(\ref{nonlocaleom}) reduces to the following local effective equation of motion in the Markovian approach.
\begin{equation}\label{Eq_conf}
    \Ddot{\theta}+\left(2\cfrac{\Dot{a}}{a}+\Gamma_{\mathrm{eff}} \right)\Dot{\theta}+a^2\cfrac{\Lambda^4}{f^2}\sin{\theta} = 0,
\end{equation}
where the dot represents the derivative with respect to the conformal time $\tau$.  As in the Minkowski case, the form of the local dissipative term in the FLRW
equation is also supported by an energy-balance argument, as shown in
Appendix~\ref{app:energy_balance_FLRW}.

Now, let us rewrite Eq.~\eqref{Eq_conf} in dimensionless variables and assume that the universe is radiation-dominated (then $a\propto \tau$ in conformal time):
\begin{equation}\label{Eq_conf_dimless}
    \theta''+ \left( \cfrac{2}{\eta}+\Gamma  \right)\theta'+\eta^2\sin{\theta} = 0,
\end{equation}
where $\Gamma=\Gamma_{\rm eff}f/\Lambda^2$ is the dimensionless damping parameter and the prime represents the derivative with respect to the dimensionless time parameter $\eta=\Lambda^2\tau/f$. We normalize the scale factor as $a(\tau)=\tau/\tau_{\rm in}$, where $\tau_{\rm in}\simeq m_\theta^{-1}=f/\Lambda^2$ denotes the onset of oscillations. In terms of the dimensionless conformal time $\eta=\tau/\tau_{\rm in}$, this gives $a(\eta)=\eta$ and
$\eta_{\rm in}=1$.

\subsection{Numerical results}
\begin{figure}[ht]
    \centering
    \begin{subfigure}{0.5\textwidth}
    \centering
        \includegraphics[width=1\linewidth]{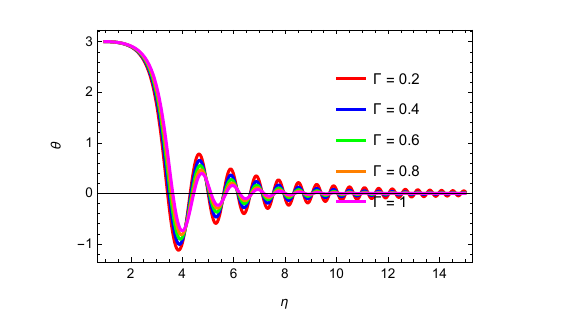}
        \caption{Numerical solutions for sample values of $\Gamma\leqslant 1$.}
        \label{fig:a1}
    \end{subfigure}%
    \begin{subfigure}{0.5\textwidth}
    \centering
        \includegraphics[width=1\linewidth]{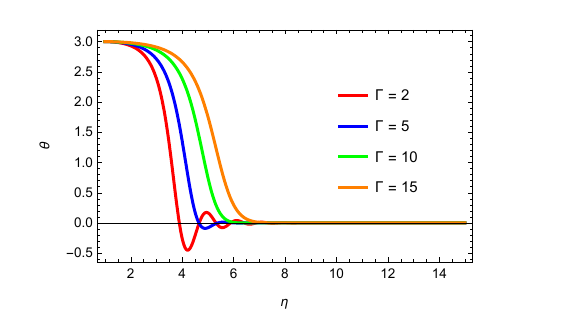}
        \caption{Numerical solutions for sample values of $\Gamma > 1$.}
        \label{fig:b1}
    \end{subfigure}
\caption{Large initial misalignment: numerical solutions of the equation of
motion in the conformal FLRW background, Eq.~\eqref{Eq_conf_dimless}, with
$\theta_{\rm in}=3.1$ and $\dot\theta_{\rm in}=0$ for different values of
$\Gamma$.}
\label{Conf1}
\end{figure}

\begin{figure}[ht]
    \centering
    \begin{subfigure}{0.5\textwidth}
    \centering
        \includegraphics[width=1\linewidth]{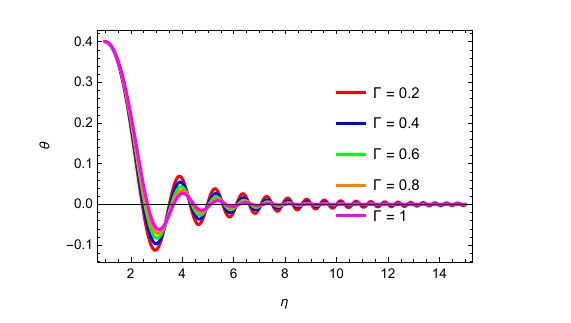}
        \caption{Numerical solutions for sample values of $\Gamma\leqslant1$.}
        \label{fig:a2}
    \end{subfigure}%
    \begin{subfigure}{0.5\textwidth}
    \centering
        \includegraphics[width=1\linewidth]{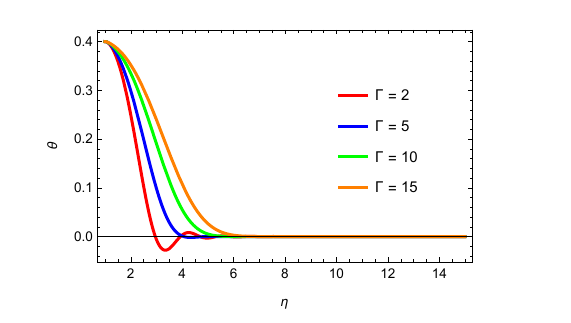}
        \caption{Numerical solutions for sample values of $\Gamma> 1$.}
        \label{fig:b2}
    \end{subfigure}
\caption{Small initial misalignment: numerical solutions of the same equation,
Eq.~\eqref{Eq_conf_dimless}, with $\theta_{\rm in}=0.4$ and
$\dot\theta_{\rm in}=0$ for different values of $\Gamma$.}
\label{Conf2}
\end{figure}

We illustrate numerical solutions of Eq.~\eqref{Eq_conf_dimless} for different
values of $\Gamma$ in Fig.~\ref{Conf1}, choosing $\theta_{\rm in}$ near $\pi$.
For comparison, Fig.~\ref{Conf2} shows the corresponding evolution for a small
initial phase, where no significant delay in the onset of oscillations is
observed. In an FLRW background, the expansion itself contributes a friction
term. This is visible in Fig.~\ref{fig:a1}, where the onset of oscillations is
delayed by approximately one Hubble time. As in the Minkowski case, larger
values of $\Gamma$ delay the onset of the motion. However, $\theta$ approaches
zero much faster than in flat spacetime.

\newpage
\section{Baryon asymmetry\label{baryonasymmetry}}
In this section, we calculate the baryon asymmetry using the solutions of the equation of motion, obtained in the previous sections. 

Following \cite{Dolgov_1997}, the baryon ($B$) and antibaryon ($\overline{B}$) number densities in case of Minkowski metric can be calculated as follows:
\begin{equation}\label{BarDensity}
    n_{B,\overline{B}}=\cfrac{g^2f^2}{2\pi^2}\int_0^\infty \omega^2 d\omega \left|\int_{-\infty}^{+\infty} e^{2i\omega t \pm i\theta(t)}dt \right|^2\;,
\end{equation}
where $+\theta(t)$ refers to baryons and $-\theta(t)$ to antibaryons. Note that $\omega$ in baryon asymmetry integrals is not the same variable as in equations of motion despite the same notation.

Let us denote the integral with respect to $t$ as follows: 
$$\int_{-\infty}^{+\infty} e^{2i\omega t \pm i\theta(t)}dt = N_\pm(\omega)\;,$$ 
then it is easy to show that:
\begin{equation}\label{N}
    N_\pm(\omega) = -\cfrac{ie^{\pm i\theta_{in}}}{2\omega}+\cfrac{i}{2\omega}+\int_{0}^{+\infty} e^{2i\omega t }(e^{\pm i\theta(t)}-1)dt\;,
\end{equation}
where we discard delta functions, due to the $\omega^2$ external integral. This can also be justified by the strict lower limit of $\omega = m_Q+m_L>0$. In addition, the last term in \eqref{N} will be calculated numerically, similar to the integral in the previous section.

\begin{figure}[ht]
    \centering
    \begin{subfigure}{0.5\textwidth}
    \centering
        \includegraphics[width=1\linewidth]{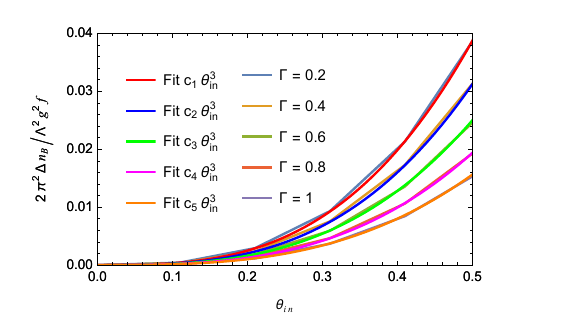}
        \caption{Numerical solutions for sample values of $\Gamma\leqslant1$.}
        \label{fig:a3}
    \end{subfigure}%
    \begin{subfigure}{0.5\textwidth}
    \centering
        \includegraphics[width=1\linewidth]{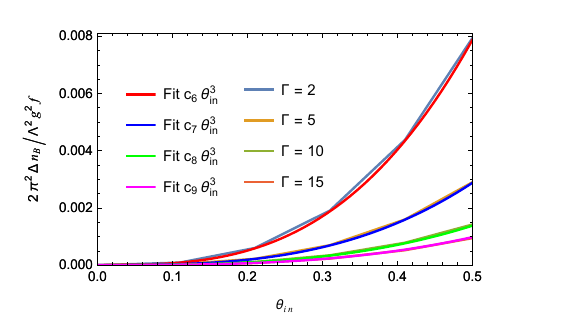}
        \caption{Numerical solutions for sample values of $\Gamma> 1$.}
        \label{fig:b3}
    \end{subfigure}
\caption{Baryon asymmetry in Minkowski metric for small initial phase for larger values of $\Gamma$ with cubic fit functions. Presented for validation of our approach. Values of $c_i$ are as follows: $c_1\approx0.31,\,c_2\approx 0.25,\,c_3\approx 0.2,\,c_4\approx 0.155,\,c_5\approx 0.125, c_6\approx0.063,\,c_7\approx 0.023,\,c_8\approx 0.011,\,c_9\approx 0.0078$.}
\label{Res1}
\end{figure}

Now, we are ready to calculate the baryon asymmetry.  First, we must confirm that our approach reproduces the results of Ref.~\cite{Dolgov_1997}, where baryon asymmetry was found to be proportional to $\theta^3_{in}$ in case of small oscillations. For this purpose, we consider small values of the initial phase and make a plot with cubic fit, displayed in Fig.~\ref{Res1}.

\begin{figure}[ht]
    \centering
    \begin{subfigure}{0.5\textwidth}
    \centering
        \includegraphics[width=1\linewidth]{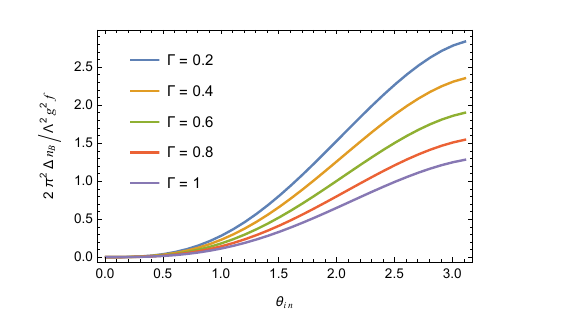}
        \caption{Numerical solutions for sample values of $\Gamma\leqslant1$.}
        \label{fig:a4}
    \end{subfigure}%
    \begin{subfigure}{0.5\textwidth}
    \centering
        \includegraphics[width=1\linewidth]{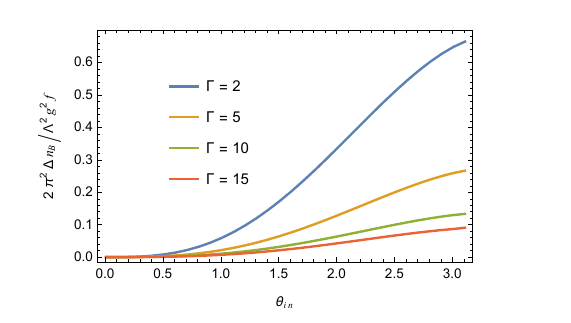}
        \caption{Numerical solutions for sample values of $\Gamma> 1$.}
        \label{fig:b4}
    \end{subfigure}
\caption{Baryon asymmetry $\Delta n_B$ for different initial phases in
Minkowski spacetime. After $\theta_{\rm in}\simeq 1$, the growth of particle
production slows considerably, and as $\theta_{\rm in}$ approaches $\pi$, it
tends toward saturation. For small values of $\Gamma$, the curve is not
significantly affected by $\Gamma$, whereas larger values of $\Gamma$ lead to
stronger suppression, as shown in panel (b).}
\label{Res2}
\end{figure}

Next, we present the plot for larger values of the initial phase. We demonstrate our result for the Minkowski metric in Fig.~\ref{Res2}.  For small oscillations, the period $T\sim 1/m_\theta$, which is not the case for a large initial phase. The apparent saturation of particle production is likely to be caused by the fact that, as the initial phase approaches $\pi$, the oscillation period becomes noticeably longer than in the harmonic approximation. Although the difference between cubic dependence and our calculations is noticeable, both values are of the same order.

 Now let us turn to the case of the conformal FLRW metric, in which we have the baryon number density as follows:
 \begin{equation}
    n(Q,\overline{L})=\cfrac{1}{V} \sum_{s_Q, s_{\overline{L}} } \int \widetilde{dp}\widetilde{dq} \bigg | \langle Q(p,s_Q),\overline{L}(q,s_{\overline{L}}) |i\cfrac{gf}{\sqrt{2}} \int d^4x  \,a(\tau)\overline{Q}(x)L(x)e^{i\theta(\tau)}|0 \rangle\bigg|^2 ,
\end{equation}
what we found to be equal to
\begin{equation}\label{Dens_exp}
    n_{B,\overline{B}}(\tau) = \cfrac{g^2 f^2}{2\pi^2 a^3(\tau)}\int_0^{\infty} \omega^2 d\omega \left| \int_{\tau_{in}}^\tau d\tau' \,a(\tau') e^{2i\omega \tau' \pm i\theta(\tau')} \right|^2.
\end{equation}

The frequency integral is formally cut off at the same scale
$\omega_{\rm cut}\sim f$ discussed in Sec.~\ref{sec:local}. Since the
time-dependent source varies on the characteristic pNGB time scale,
frequencies well above $m_\theta$ are expected to give mainly rapidly
oscillating contributions. We have verified numerically that raising the upper
limit beyond a few times $m_\theta$ changes the resulting asymmetry only
negligibly, so that the results are insensitive to the precise value of the
cutoff within the hierarchy $m_\theta\ll\omega_{\rm cut}$.

\begin{figure}[ht]
    \centering
    \begin{subfigure}{0.5\textwidth}
    \centering
        \includegraphics[width=1\linewidth]{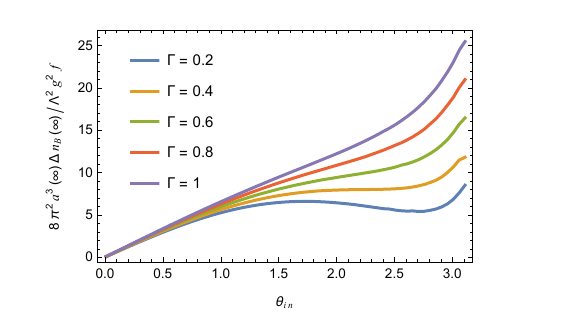}
        \caption{Numerical solutions for sample values of $\Gamma\leqslant1$.}
        \label{fig:a5}
    \end{subfigure}%
    \begin{subfigure}{0.5\textwidth}
    \centering
        \includegraphics[width=1\linewidth]{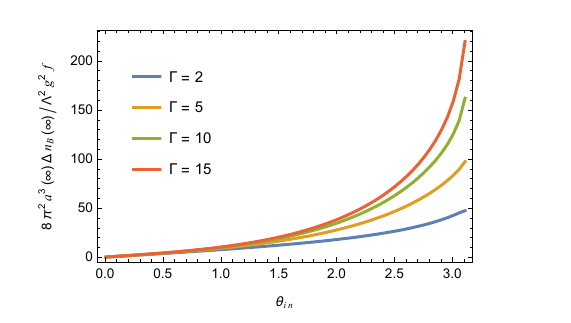}
        \caption{Numerical solutions for sample values of $\Gamma> 1$.}
        \label{fig:b5}
    \end{subfigure}
\caption{Baryon asymmetry $a^3(\infty)\Delta n_{B}(\infty)$ for given values
of initial phase in the conformal FLRW spacetime. The quantity on the vertical axis is denoted $\kappa(\theta_{\rm in})$ in the text. We may see that the behavior of the curve is affected significantly by the value of $\Gamma$. For small values of $\Gamma$ (Fig.~\ref{fig:a5}) we
observe the decrease in asymmetry for large initial phase. For small initial
phases the asymmetry is proportional to $\theta_{in}$. For
$\Gamma \gtrsim 1$ (Fig.~\ref{fig:b5}) we see nearly exponential dependence of
asymmetry on initial phase as phase approaches $\pi$.}

\label{AsymExp}
\end{figure}

Using the expression \eqref{Dens_exp}, we calculate
$a^3(\infty)\Delta n_{B}(\infty)$, which approaches a constant
late-time value and can be compared with the corresponding result in
Minkowski spacetime. We illustrate
$a^3(\infty)\Delta n_{B}(\infty)$ as a function of the initial
phase in Fig.~\ref{AsymExp}. In practice, the late-time limit can be
approximated by $\tau_{\rm end}\simeq 10f/\Lambda^2$, as can be seen
from Figs.~\ref{Conf1} and~\ref{Conf2}. We also set the normalization
$a(\tau_{\rm in})=1$ and identify the lower limit $\tau_{\rm in}$ of
the production integral with the onset of the relevant $\theta$ evolution,
$\tau_{\rm in}\simeq \tau_{\rm osc}$. Since the expansion is assumed
to be adiabatic, $a^3 s$ is conserved, leading to
$a^3(\tau_{\rm end})s(T_{\rm end})\simeq s(T_{\rm osc})$.

Consequently, the ratio $\Delta n_B/s$ may be evaluated using the
comoving baryon number at $\tau_{\rm end}$ and the entropy density at
$T_{\rm osc}$. With the help of Fig.~\ref{AsymExp}, the baryon-to-entropy-density
ratio is estimated as
\begin{equation}
\label{baryontoentropy}
\frac{\Delta n_{\mathrm{B}}(\tau_{\rm end})}{s(T_{\rm end})}
\simeq
\frac{\Delta n_{\mathrm{B}}(\tau_{\rm end})a^3(\tau_{\rm end})}{s(T_{\rm osc})}
\simeq
0.25\,
\frac{g^{2}}{g_{*}^{1/4}}
\frac{f}{m_{\mathrm{Pl}}}
\left(\frac{\Lambda}{m_{\mathrm{Pl}}}\right)^{2}
\left(\frac{H_{\mathrm{osc}}}{m_{\mathrm{Pl}}}\right)^{-3/2}
\frac{\kappa(\theta_{\rm in})}{4} .
\end{equation}
Here, $m_{\mathrm{Pl}}=1.22\times 10^{19}\,{\rm GeV}$ is the Planck
energy and $H_{\mathrm{osc}}$ is the Hubble parameter when the oscillations
begin. Since this occurs after inflation, we have $H_{\mathrm{osc}}<H_{\star}$,
where $H_{\star}$ is the Hubble
parameter during inflation. The quantity $\kappa(\theta_{\rm in})$ is extracted
from the comoving asymmetry $a^3(\tau_{\rm end})\Delta n_B(\tau_{\rm end})$;
namely, it corresponds to the $y$-axis values in Fig.~\ref{AsymExp}. The
entropy density is given by
$s(T_{\mathrm{osc}})=\frac{2\pi^2}{45}g_{*s}T_{\mathrm{osc}}^3$, where
the corresponding temperature is
\begin{equation}
T_{\mathrm{osc}}
=
\left(\frac{90}{8\pi^3 g_{*}}\right)^{1/4}
\sqrt{H_{\mathrm{osc}}m_{\mathrm{Pl}}}\, .
\end{equation}

Note that the effective degrees of freedom associated with the entropy and the energy densities, $g_{* s}$ and $g_{*}$ respectively, are approximately equal at high temperature; for temperatures above the SM value, we have $g_{* s}\simeq g_{*}=106.75$. This value could change slightly if there are new degrees of freedom, but the effect of this change in Eq.~(\ref{baryontoentropy}) is clearly minor. Assuming that the expansion of the universe is adiabatic, in particular in the absence of significant entropy production after the reheating, the entropy and baryon number are conserved in the co-moving volume; therefore, the estimation in Eq.~(\ref{baryontoentropy}) is valid for the present universe, $(\Delta n_{\mathrm{B}}/s)_0\simeq \Delta n_{\mathrm{B}}/s$. 

Two constraints must be respected in what follows: $g\ll1$, which guarantees
a perturbative coupling, and $\Lambda\ll f$, which follows from the way the
problem was set up.
Furthermore, the field $\theta$ must remain a light spectator during
inflation, $H_\star \gg m_\theta = \Lambda^2/f \simeq H_{\rm osc}$, with
$H_\star$ the inflationary Hubble parameter. When the damping rate exceeds the expansion rate at the onset of oscillations, $\Gamma_{\rm eff}\gg H_{\rm osc}$, or equivalently $\Gamma\gg1$
in terms of the dimensionless damping parameter, the effects of the expansion
are not important, and the Minkowski analysis is adequate.
Otherwise the expansion must be taken into account, as we do here.

Constraining the parameter space is beyond the scope of this paper, but
it is instructive to examine a few representative parameter sets, collected in
Table~\ref{tab:benchmarks} and evaluated at $\bar\theta_{\rm in}=0.95\pi$.
All four reproduce the observed asymmetry
$(\Delta n_{\rm B}/s)_0\simeq8.6\times10^{-11}$ through
Eq.~(\ref{baryontoentropy}), with $\kappa(\bar\theta_{\rm in})$ read from
Fig.~\ref{AsymExp} for the corresponding value of $\Gamma$.
For each set, the requirement that $\theta$ remain a light spectator during
inflation, $H_\star\gg m_\theta=\Lambda^2/f$, provides a lower limit on the
inflationary Hubble parameter, whereas the upper limit follows from the baryon isocurvature constraint discussed in the next section. For all sets considered here, this constraint is more restrictive than both the Planck bound $H_\star\lesssim6.1\times10^{13}$~GeV~\cite{Planck:2018jri}
and the requirement $H_\star<f$. The reheating temperature is bounded from below by $T_{\rm osc}$ and from above by the corresponding limit on $H_\star$. The first three sets admit a range of inflationary scales spanning up to
three orders of magnitude and are therefore viable. By contrast, the fourth
set has a considerably milder hierarchy, $f/\Lambda\simeq23$, and hence a
larger $m_\theta$. Its isocurvature limit lies below the spectator requirement
$H_\star\gg m_\theta$, so it is excluded despite reproducing the observed
asymmetry.

\begin{table}[t]
\centering
\begin{tabular}{lcccc}
\hline\hline
 & Set 1 & Set 2 & Set 3 & Set 4 \\
\hline
\noalign{\vskip 5pt}
$\Gamma$                                  & $1$   & $0.2$ & $5$   & $1$   \\
\noalign{\vskip 5pt}
$\kappa(\bar\theta_{\rm in})$                         & $22.5$  & $6.5$  & $80.8$  & $22.5$  \\
\noalign{\vskip 5pt}
$g$                                       & $10^{-2}$ & $10^{-3}$ & $10^{-3}$ & $2\times10^{-2}$ \\
\noalign{\vskip 5pt}
$f$ [GeV]                                 & $1.1\times10^{13}$ & $1.7\times10^{15}$
                                          & $1.4\times10^{14}$ & $9.5\times10^{13}$ \\
                                          \noalign{\vskip 5pt}
$\Lambda$ [GeV]                           & $4.8\times10^{9}$  & $4.1\times10^{12}$
                                          & $9.9\times10^{10}$ & $4.2\times10^{12}$ \\
                                          \noalign{\vskip 5pt}
$f/\Lambda$                               & $2.3\times10^{3}$  & $4.2\times10^{2}$
                                          & $1.4\times10^{3}$  & $23$ \\
                                          \noalign{\vskip 5pt}
\hline
\noalign{\vskip 5pt}
$H_{\rm osc}\simeq m_\theta$ [GeV]               & $2.1\times10^{6}$  & $9.9\times10^{9}$
                                          & $7.0\times10^{7}$  & $1.9\times10^{11}$ \\
                                          \noalign{\vskip 5pt}
$T_{\rm osc}$ [GeV]                       & $1.2\times10^{12}$ & $8.4\times10^{13}$
                                          & $7.1\times10^{12}$ & $3.6\times10^{14}$ \\
                                          \noalign{\vskip 5pt}
\hline
\noalign{\vskip 5pt}
$R(\bar\theta_{\rm in})$                         & $1.0$ & $1.6$ & $1.3$ & $1.0$ \\
\noalign{\vskip 5pt}
\makecell[l]{$H_\star$ [GeV]\\[2pt]
  \small $(m_\theta \ll H_\star \lesssim H_\star^{\rm iso})$}
    & $2.1\times10^{6}$--$3.3\times10^{9}$
    & $9.9\times10^{9}$--$3.2\times10^{11}$
    & $7.0\times10^{7}$--$3.2\times10^{10}$ & --- \\
    \noalign{\vskip 5pt}
\makecell[l]{$T_{\rm rh}$ [GeV]\\[2pt]
  \small $(T_{\rm osc} \ll T_{\rm rh} \lesssim T_{\rm rh}^{\rm max})$}
    & $1.2\times10^{12}$--$4.9\times10^{13}$
    & $8.4\times10^{13}$--$4.8\times10^{14}$
    & $7.1\times10^{12}$--$1.5\times10^{14}$ & --- \\
    \noalign{\vskip 5pt}
                             \hline
Status                                    & viable & viable & viable & not viable \\
\hline\hline
\end{tabular}
\caption{Sample parameter sets, all evaluated at
$\bar\theta_{\rm in}=0.95\pi$ and reproducing the observed asymmetry
$(\Delta n_{\rm B}/s)_0\simeq8.6\times10^{-11}$. Here $H_\star^{\rm iso}$ denotes the upper limit from the isocurvature constraint, Eq.~(\ref{Hstar-iso-bound}), whereas the lower limit is the
spectator requirement $H_\star\gg m_\theta$. The corresponding limit on the
reheating temperature, $T_{\rm rh}^{\rm max}$, follows from
$H_\star^{\rm iso}$ assuming instantaneous reheating. For Set~4, the isocurvature limit lies below $m_\theta$, so no admissible inflationary scale
exists.}
\label{tab:benchmarks}
\end{table}

\section{Isocurvature bounds}
\label{sec:iso}

The gauge-invariant baryon isocurvature perturbation is defined
as~\cite{Lyth:2009zz}
\begin{equation}
S_{B\gamma}
\equiv
\frac{\delta n_B}{n_B}
-\frac{3}{4}\frac{\delta\rho_\gamma}{\rho_\gamma}.
\end{equation}
In the setup considered here, the radiation bath originates from inflaton
reheating, whereas $\theta$ is an energetically negligible spectator field.
Its fluctuations therefore do not source the curvature perturbation and are
uncorrelated with the inflaton-generated adiabatic mode. Instead, they perturb
the final baryon number through the dependence of the produced asymmetry on
the initial phase $\theta_{\rm in}$. We therefore choose uniform-radiation hypersurfaces, on which $\delta\rho_\gamma=0$. Patches with slightly different $\theta_{\rm in}$ then
acquire different values of $n_B$ at the same photon density. Thus, in this spectator limit, the baryon isocurvature perturbation
reduces to
\begin{equation}
\label{fractional-pert}
S_{B\gamma}
=
\frac{\delta n_B}{n_B}.
\end{equation}
Since $s\propto \rho_\gamma^{3/4}$ for a radiation bath, the entropy density is
also unperturbed on these hypersurfaces, so equivalently
$S_{B\gamma}=\delta\ln(n_B/s)$.

In the present mechanism, the late-time comoving asymmetry in each
super-horizon patch is a function of the initial angle of that patch,
\begin{equation}
\label{eq:F}
a^3 n_B(\infty) \equiv F(\theta_{\rm in}) .
\end{equation}
The only source of baryon-number fluctuations from patch to patch is therefore
the variation of $\theta_{\rm in}$ itself.

Let us write the initial angle in a given patch as
\begin{equation}
\theta_{\rm in}(\mathbf{x})
=
\bar\theta_{\rm in}+\delta\theta(\mathbf{x}),
\end{equation}
where $\delta\theta$ is the super-horizon phase fluctuation frozen since
horizon exit. Expanding Eq.~\eqref{eq:F} to linear order, the fractional baryon-number
perturbation, $S_{B\gamma}=\delta n_B/n_B$, becomes
\begin{equation}
\label{eq:logarithmic-derivative}
S_{B\gamma}
=
\frac{F(\bar\theta_{\rm in}+\delta\theta)-F(\bar\theta_{\rm in})}
{F(\bar\theta_{\rm in})}
\simeq
\left.
\frac{d \ln F(\theta_{\rm in})}{d \theta_{\rm in}}
\right|_{\bar\theta_{\rm in}}
\delta\theta .
\end{equation}
Thus, the sensitivity of the asymmetry to the initial condition is controlled
by the logarithmic derivative of the late-time comoving asymmetry with respect
to $\theta_{\rm in}$.

As stated above, the angular field is a light spectator during inflation,
$m_\theta=\Lambda^2/f\ll H_\star$, by construction. A light scalar field in
quasi--de~Sitter space acquires super-horizon fluctuations with an
approximately scale-invariant spectrum. The amplitude per logarithmic interval
of $k$ at horizon exit is
\begin{equation}
\mathcal{P}_{\delta\phi}^{1/2}(k)
=
\frac{H_\star}{2\pi},
\qquad\text{hence}\qquad
\mathcal{P}_{\delta\theta}^{1/2}(k)
=
\frac{\mathcal{P}_{\delta\phi}^{1/2}(k)}{f}
=
\frac{H_\star}{2\pi f},
\end{equation}
where $\phi=f\theta$ is the canonically normalized field. The linear relation in Eq.~(\ref{eq:logarithmic-derivative}) then gives the
corresponding isocurvature power-spectrum amplitude,\footnote{Since $S_{B\gamma}=T\,\delta\theta$ at linear order, with
$T=d\ln F/d\theta_{\rm in}$, the two-point functions are related by
$\langle S_{B\gamma}S_{B\gamma}\rangle
=T^2\langle\delta\theta\,\delta\theta\rangle$, or equivalently
$\mathcal{P}_{B\gamma}(k)=T^2\mathcal{P}_{\delta\theta}(k)$ in terms of
dimensionless power spectra.}
\begin{equation}
\label{eq:isoamplitude}
\mathcal{P}_{B\gamma}^{1/2}(k)
=
\left|
\frac{d \ln F}{d \theta_{\rm in}}
\right|
\frac{H_\star}{2\pi f},
\end{equation}
with the derivative evaluated at the background value of $\theta_{\rm in}$ in
the patch under consideration. Thus, in the linear approximation, the induced baryon isocurvature spectrum inherits the approximate scale invariance of the spectator-field fluctuations. The expression above assumes that the fluctuation imprinted at horizon exit remains essentially unchanged until the onset of
oscillations. This assumption holds while $m_\theta\ll H$, since the field is
overdamped. The background value and the fluctuation are therefore frozen from
horizon exit through reheating and into radiation domination.

\begin{figure}[ht]
    \centering
    \begin{subfigure}{0.6\textwidth}
    \centering
        \includegraphics[width=1\linewidth]{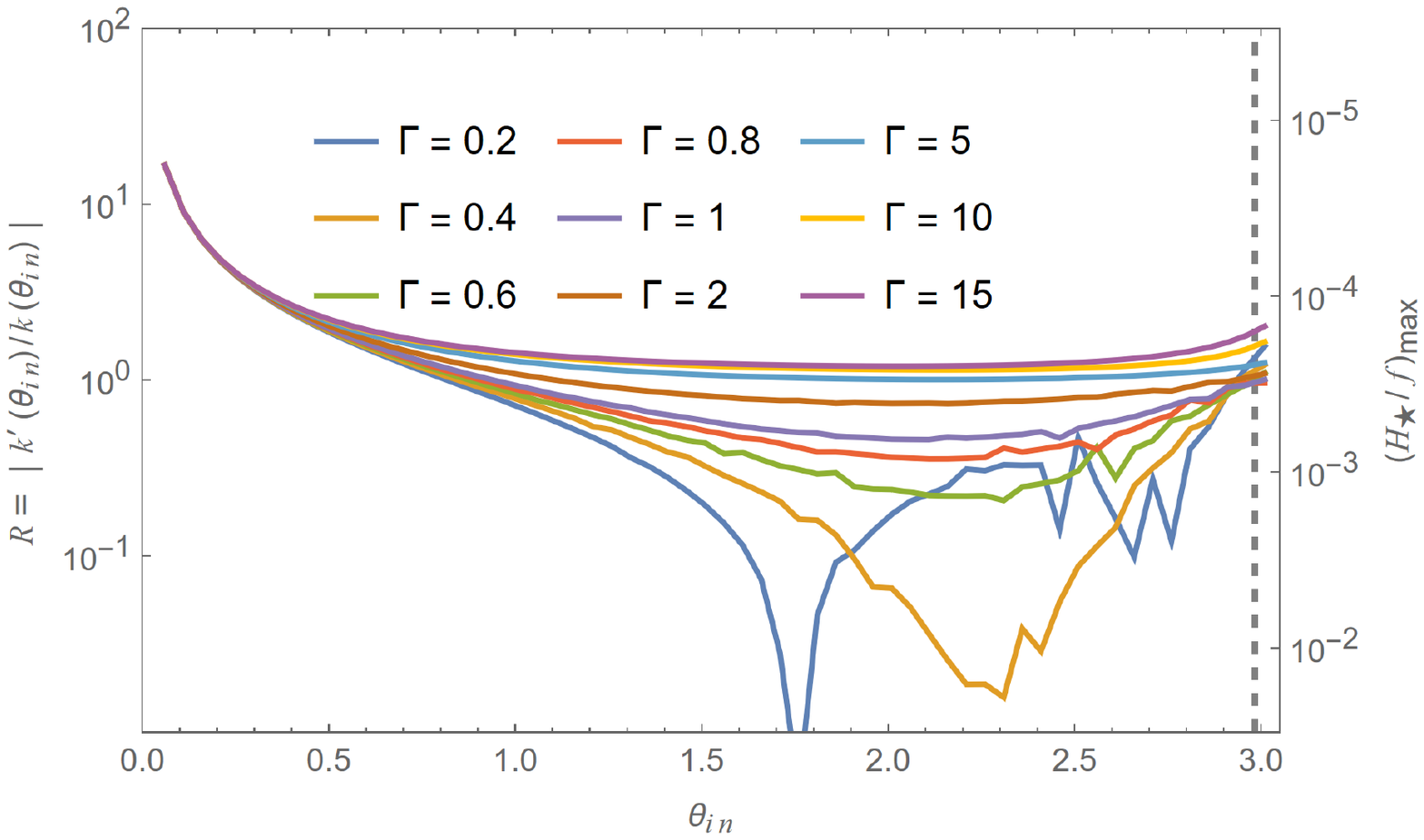}
    \end{subfigure}
\caption{The logarithmic derivative $R(\theta_{\rm in})$ entering the
isocurvature power spectrum in Eq.~(\ref{eq:iso-power}), obtained by
differentiating the numerical results shown in Fig.~\ref{AsymExp}. The small oscillations at intermediate $\theta_{\rm in}$ mainly reflect numerical differentiation, whereas the pronounced dips for $\Gamma\lesssim0.4$ correspond to stationary points of $\kappa(\theta_{\rm in})$. The right-hand axis shows the corresponding upper bound on $H_\star/f$ from Eq.~(\ref{Hstar-iso-bound}). The vertical dashed line denotes $\bar\theta_{\rm in}=0.95\pi$.}
\label{logarithmic-derivative}
\end{figure}

The baryon-to-entropy ratio obtained in Sec.~\ref{baryonasymmetry} takes the
form
\begin{equation}
\frac{\Delta n_B(\tau_{\rm end})}{s(T_{\rm end})}
\propto
\kappa(\theta_{\rm in}),
\end{equation}
as in Eq.~(\ref{baryontoentropy}). Here, $\kappa(\theta_{\rm in})$ is the dimensionless late-time comoving asymmetry extracted from the numerical solutions, shown in Fig.~\ref{AsymExp} as a function of $\theta_{\rm in}$ for each value of the damping parameter $\Gamma$. Since $F(\theta_{\rm in})$ and $\kappa(\theta_{\rm in})$ differ only by constant factors, namely powers of $\Lambda$, $g$, and $f$ together with numerical constants, the two have the same logarithmic derivative,
$d\ln F/d\theta_{\rm in}=d\ln\kappa/d\theta_{\rm in}$. The isocurvature power spectrum is therefore
\begin{equation}
\label{eq:iso-power}
\mathcal{P}_{B\gamma}(k)
=
R^2(\theta_{\rm in})
\left(\frac{H_\star}{2\pi f}\right)^2,
\qquad
R(\theta_{\rm in})
\equiv
\left|
\frac{d \ln \kappa(\theta_{\rm in})}{d \theta_{\rm in}}
\right| ,
\end{equation}
where $R(\theta_{\rm in})$, obtained from Fig.~\ref{AsymExp}, is displayed in
Fig.~\ref{logarithmic-derivative}.

The current bound\footnote{Since the CMB responds at linear order only to the
total matter entropy perturbation, a baryon isocurvature mode is
observationally equivalent to a cold dark matter isocurvature mode of amplitude
$S_{\rm eff}=(\Omega_B/\Omega_{\rm CDM})S_{B\gamma}$. More generally, a scalar
that survives to the present and contributes to the dark matter would add a
term $(\Omega_\theta/\Omega_{\rm CDM})S_{\theta\gamma}$ to the effective
isocurvature. In the present model, $\theta$ converts into baryons and does
not survive, so this contribution is absent. The constrained isocurvature
power is therefore
$\mathcal{P}_{\rm eff}=(\Omega_B/\Omega_{\rm CDM})^2\mathcal{P}_{B\gamma}$~\cite{Inomata:2018htm},
which corresponds to $\mathcal{P}_{II}$ in the notation of
Ref.~\cite{Planck:2018jri}. We use the Planck
uncorrelated, scale-invariant isocurvature bound for the ``axion~I'' case,
reported in Sect.~9.1 and Table~14 of Ref.~\cite{Planck:2018jri}. This provides the appropriate cold-dark-matter isocurvature (CDI) template since the mode generated by $\delta\theta$ is uncorrelated with the adiabatic perturbation and approximately scale invariant. The bound is
$\beta_{\rm iso}\lesssim0.038$ at
$k_\star=0.05\ {\rm Mpc}^{-1}$, where the primordial isocurvature fraction
is $\beta_{\rm iso}(k)=\mathcal{P}_{II}(k)/
[\mathcal{P}_{\mathcal{RR}}(k)+\mathcal{P}_{II}(k)]$. Here,
$\mathcal{P}_{\mathcal{RR}}$ is the curvature power spectrum, also commonly
denoted $\mathcal{P}_{\zeta}$. Together with
$\mathcal{P}_{\mathcal{RR}}=A_s\simeq2.1\times10^{-9}$ at the same pivot
scale $k_\star=0.05\ {\rm Mpc}^{-1}$ and
$\Omega_B/\Omega_{\rm CDM}\simeq0.187$, from Table~2 of
Ref.~\cite{Planck:2018vyg}, this yields
$\mathcal{P}_{B\gamma}\lesssim
(\Omega_{\rm CDM}/\Omega_B)^2(\beta_{\rm iso}\mathcal{P}_{\mathcal{RR}})
\simeq2.3\times10^{-9}$.} is
\begin{equation}
\mathcal{P}_{B\gamma}(k_\star)
\lesssim
2.3\times10^{-9},
\qquad
k_\star=0.05\ {\rm Mpc}^{-1}.
\end{equation}
Using Eq.~(\ref{eq:iso-power}), this translates into
\begin{equation}
\label{Hstar-iso-bound}
\frac{H_\star}{f}
\lesssim
\frac{3.0\times 10^{-4}}{R(\theta_{\rm in})}.
\end{equation}

As Fig.~\ref{logarithmic-derivative} shows, $R(\theta_{\rm in})$ takes values close to unity at the large angles of interest here, whereas it becomes very small near the stationary points of $\kappa(\theta_{\rm in})$, where the linear isocurvature is suppressed. At small angles, $R$ grows as $\theta_{\rm in}^{-1}$, so the bound on $H_\star/f$ tightens correspondingly. Nevertheless, the constraint is readily satisfied over broad regions of parameter space. The sample sets in Table~\ref{tab:benchmarks} illustrate this:
three of the four satisfy Eq.~(\ref{Hstar-iso-bound}) with an allowed inflationary window spanning up to three orders of magnitude, whereas the fourth is excluded by the isocurvature bound despite reproducing the observed asymmetry. Thus, the isocurvature constraint restricts the ratio $H_\star/f$, but leaves viable large-misalignment parameter space for the benchmark ranges considered
here.

It is instructive to compare this conclusion with Appendix~A of Ref.~\cite{Simone_2016}, where baryogenesis from the decay of the oscillating condensate is estimated to be too small once cosmological expansion and isocurvature constraints are included. Their estimate starts from the small-amplitude flat-space result of Refs.~\cite{Dolgov_1995,Dolgov_1997},
where the generated asymmetry scales cubically with the oscillation amplitude. The effect of expansion is incorporated by allowing the oscillation amplitude to redshift until the time of decay, $H=\Gamma_\phi$, with $\Gamma_\phi=\beta m_\phi^3/f^2$. Thus, expansion enters through a redshift estimate for the oscillation amplitude, rather than through an explicit solution of the expanding-background equation of motion. In radiation domination, this gives a suppression factor $(\Gamma_\phi/H_{\rm osc})^{3/4}$ for each power of the oscillation amplitude. This effect, and hence the conclusion drawn from it, applies to the regime in which the decay rate is parametrically smaller than the oscillation scale. In the present analysis, on the other hand, arbitrary amplitudes are treated directly in the expanding background, so no such factor is introduced separately. The expansion enters the equation of motion, Eq.~(\ref{Eq_conf_dimless}), and the production integral, Eq.~(\ref{Dens_exp}), directly. The competition between Hubble friction and the local damping is therefore contained in the numerical solution, and the suppression produced by expansion is already included in $\kappa(\theta_{\rm in})$.

In their parametrization, $\Gamma_\phi/H_{\rm osc}\simeq\beta\,m_\phi^2/f^2$ is suppressed by the hierarchy $m_\phi\ll f$ unless $\beta$ is correspondingly much larger than unity. Ref.~\cite{Simone_2016} takes $\beta\lesssim1$. The present model, by contrast, is not described by this late-decay parametrization. As discussed below, translating our conversion rate into their notation gives an effective $\beta$ far larger than unity throughout the range of $\Gamma$ considered here. Recovering $\beta\lesssim1$ would require $\Gamma\lesssim(m_\theta/f)^2$, which lies far below the range in which the mechanism produces the observed asymmetry.
\\

The smallness of $\beta$ in Ref.~\cite{Simone_2016} reflects the anomalous
decay modes $\phi\to WW,ZZ$, which dominate once the baryon-violating
interactions decouple. In their decay-rate estimate, the authors use
$\Gamma_\phi=\beta m_\phi^3/f^2$ as a parametrization of possible decay
channels, with the model dependence encoded in $\beta$. They also note that additional channels may be present if further sources of baryon violation contribute to $\partial_\mu j^\mu$. The model studied here is of this type. In the present model, these anomalous electroweak modes are absent from the $Q$--$L$ sector, since $Q$ and $L$ are Standard-Model singlets. Instead, the divergence of the current coupled to the pNGB is generated by the Yukawa interaction itself, as in Eq.~(\ref{currents}). The relevant process is therefore particle production from a time-dependent background rather than a late perturbative decay, namely the production of $Q$ and $L$ during the coherent pNGB evolution, with the backreaction encoded in the local coefficient $\Gamma_{\rm eff}$ rather than fixed by $m_\theta^3/f^2$. The bound in their Appendix~A is derived for $\beta\lesssim1$ and therefore does not cover a conversion of this type.

The regime relevant in the present analysis can be characterized by the ratio of the local conversion rate to the Hubble rate at the onset of oscillations. With $\Gamma_{\rm eff}=\Gamma m_\theta$ and $H_{\rm osc}\simeq m_\theta$, this ratio is the dimensionless damping parameter itself, $\Gamma_{\rm eff}/H_{\rm osc}\simeq\Gamma$, which is not parametrically small over the range $\Gamma=0.2$--$15$ explored here. This picture is visible in the numerical solutions in Fig.~\ref{Conf1}: for $\Gamma>1$ the motion is strongly damped and the field has settled by $\eta\simeq7$, whereas for $\Gamma\le1$ it oscillates with a rapidly decreasing amplitude that is negligible beyond $\eta\simeq10$. There is therefore no parametrically long redshift epoch of the type assumed in Appendix~A of Ref.~\cite{Simone_2016}. For the range of $\Gamma$ explored here, our effective conversion rate corresponds in their notation to $\beta_{\rm eff}=\Gamma f^2/m_\theta^2\gg1$, since $m_\theta=\Lambda^2/f\ll f$. For sufficiently small $\Gamma$, the conversion would become too inefficient
to account for the observed baryon asymmetry, but this does not affect the
benchmark sets identified here.

The dependence on the conversion efficiency is visible in our results. The comparison with the redshift-suppression estimate of
Ref.~\cite{Simone_2016} is meaningful only in the weakly damped regime,
$\Gamma_{\rm eff}/H_{\rm osc}\simeq\Gamma\lesssim1$. This is analogous to
$\Gamma_\phi/H_{\rm osc}\lesssim1$ in their late-decay estimate. At
$\bar\theta_{\rm in}=0.95\pi$, Fig.~\ref{fig:a5} gives a decrease from
$\kappa\simeq22.5$ at $\Gamma=1$ to $\kappa\simeq6.5$ at $\Gamma=0.2$, a
factor of a few rather than many orders of magnitude. 

\section{Discussion and Conclusion}
We have revisited the scenario of spontaneous baryogenesis by a Nambu-Goldstone boson. In the literature, the small-angle assumption has been commonly employed in approximating the cosine potential. However, the probability distribution of the phase, determined by quantum fluctuations during the inflationary stage, leads to a non-negligible probability for a noticeable phase change during inflation. This raises the question of the reliability of this approximation and necessitates the study of a possible large misalignment. 
 
Our purpose in this paper was to examine the leading effects of going beyond
the small-angle approximation commonly used in the literature, rather than to
identify a specific realization of the observed baryon asymmetry. This is
generally a difficult problem because the fermionic backreaction is nonlocal in time.
However, as we have shown, the adiabaticity conditions follow from the
hierarchy $\Lambda\ll f$ already built into the model, so that the local
Markovian description is parametrically justified. Large-angle effects are still retained through the nonlinear dependence of the potential term on $\theta$, while the resulting equation of motion remains tractable. In this local description, the damping rate is treated as a phenomenological parameter in the effective theory. Our aim is not to determine this coefficient, but to study how the generated asymmetry depends on it while focusing on the large-angle dynamics.

As a first approximation, we considered Minkowski spacetime, where the expansion
of the universe can be neglected. This limit is appropriate when the damping
rate of the pNGB oscillations is much larger than the expansion rate of the universe.  We have calculated the baryon asymmetry in the case of an initial phase close to $\pi$ since this effectively represents the largest phase value due to the local maximum at $\pi$ in the cosine potential.
Our analysis, as seen in Fig.~\ref{Res2}, indicates that the effects of a large misalignment angle are not significant in generating the required baryon asymmetry compared to the small-angle approximation in Minkowski metric.

We also present our result for the conformal FLRW metric, which is shown in Fig.~\ref{AsymExp}. We observe that in this case the asymmetry is proportional to $\theta_{in}$ in the case of a small phase and the behavior of the curve is significantly affected by $\Gamma$. For small $\Gamma$, the asymmetry can decrease at large initial phase, whereas
for $\Gamma\gtrsim1$ it shows a nearly exponential dependence on the initial
phase as the phase approaches $\pi$.

Finally, we examined the baryon-isocurvature bounds arising from inflationary
fluctuations of the initial phase. Since the pNGB is a spectator during inflation, its fluctuations do not source the primordial curvature perturbation. Instead, they modulate the final baryon
asymmetry from patch to patch, generating a baryon-isocurvature mode that is
uncorrelated with the inflaton-generated adiabatic perturbation.
The resulting bound is controlled by the logarithmic sensitivity of the final
asymmetry to the initial angle and imposes an upper limit on the ratio
$H_\star/f$. In the large-misalignment regime, the isocurvature constraint is nontrivial
and can exclude parameter choices that reproduce the observed baryon
asymmetry. Nevertheless, the benchmarks presented satisfy the bound over a range of inflationary scales.

\par Let us conclude with several remarks. The mechanism of spontaneous baryogenesis can lead to phase fluctuations that cross $\pi$. Such a crossing leads to the formation of closed domain walls, which can collapse in black holes or form baby universes creating gravitational wave background \cite{Sakharov_2021,article}. The other side of such crossing $\pi$ is the change of the sign of baryon excess and creation of macroscopic antibaryon domains in the baryon-asymmetric Universe. Even if fluctuations do not cross $\pi$, the demonstrated strong dependence of the baryon density on the value of phase leads to prediction of strong baryon density fluctuations, which can be important for early galaxy formation, favored by the JWST observations. It makes the complex of multi-messenger cosmological signatures of axion-like particle physics an exciting field for future research.

\section*{Acknowledgments}
The work of M. K. was performed in Southern Federal University with financial support of grant of Russian Science Foundation № 25-07-IF. The work of U. A. is funded by The Scientific and Technological Research Council of T\"urkiye (T\"UB\.ITAK) B\.{I}DEB 2232-A program under project № 121C067.

\appendix
\section{Energy-balance argument for the dissipative term}
\label{app:energy_balance}

In this appendix we give a simple energy-balance argument for the form of the
leading local dissipative term in the effective equation of motion.
The argument is presented first in Minkowski spacetime and then in the
conformal FLRW background.
It fixes the structure of the term, however it does not determine its
coefficient.

\subsection{In Minkowski}
\label{app:energy_balance_flat}
The form of the local dissipative term can also be inferred from energy
balance considerations. The energy density of the homogeneous pNGB field
associated with the cosine potential is
\begin{equation}
\rho_\theta =
\frac{1}{2}f^2\dot\theta^2+\Lambda^4(1-\cos\theta),
\end{equation}
and its time derivative is
\begin{equation}
\dot\rho_\theta
=
f^2\dot\theta
\left(
\ddot\theta+\frac{\Lambda^4}{f^2}\sin\theta
\right).
\end{equation}
The factor in parentheses vanishes in the absence of backreaction, in which
case the pNGB field conserves energy. Particle production provides the
channel through which energy is transferred from this coherent mode to the
fermionic sector. This loss can be written as
\begin{equation}
\dot\rho_\theta
=
-\mathcal Q_{\rm diss},
\qquad
\mathcal Q_{\rm diss}\geq0 ,
\end{equation}
where $\mathcal Q_{\rm diss}$ denotes the coarse-grained dissipative transfer
rate. Then, at the level of the local equation of motion, we have
\begin{equation}
\ddot\theta+\frac{\Lambda^4}{f^2}\sin\theta
=
-R_{\rm diss}
\end{equation}
with
\begin{equation}
\mathcal Q_{\rm diss}
=
f^2\dot\theta\,R_{\rm diss}.
\end{equation}
Since $\mathcal Q_{\rm diss}\geq0$, the dissipative backreaction term must satisfy
\begin{equation}
\dot\theta\,R_{\rm diss}\geq0 .
\end{equation}
Dissipative energy loss should occur for either sign of $\dot\theta$, so this
condition requires $R_{\rm diss}$ to change sign with $\dot\theta$. The adiabatic expansion suppresses terms with additional time derivatives.
Thus, close to $\dot\theta=0$, the dissipative backreaction takes the analytic
form
\begin{equation}
R_{\rm diss}
=
\Gamma_{\rm eff}\dot\theta
+
O(\dot\theta^3),
\qquad
\Gamma_{\rm eff}\geq0 .
\end{equation}
Keeping only the leading adiabatic contribution yields the local friction
term $\Gamma_{\rm eff}\dot\theta$, which in the dimensionless variables of
the main text corresponds to the damping parameter
$\Gamma=\Gamma_{\rm eff}f/\Lambda^2$ in Eq.~(\ref{Minkowskieom}).

\subsection{In conformal FLRW background}
\label{app:energy_balance_FLRW}

The same reasoning in the previous section applies in an expanding background, with the expansion
terms included in the energy balance. The physical energy density and
pressure of the homogeneous pNGB field are
\begin{equation}
\rho_\theta
=
\frac{f^2}{2a^2}\dot\theta^2
+
\Lambda^4(1-\cos\theta),
\qquad
p_\theta
=
\frac{f^2}{2a^2}\dot\theta^2
-
\Lambda^4(1-\cos\theta),
\end{equation}
where the dot denotes differentiation with respect to conformal time $\tau$.

In the absence of energy transfer to other degrees of freedom, the scalar
energy density obeys
\begin{equation}
\dot\rho_\theta
+
3\mathcal H(\rho_\theta+p_\theta)
=
0 .
\end{equation}
The dissipative transfer of energy from this coherent mode to the
fermionic sector modifies the scalar energy balance as
\begin{equation}
\dot\rho_\theta
+
3\mathcal H(\rho_\theta+p_\theta)
=
-\mathcal Q_{\rm diss},
\qquad
\mathcal Q_{\rm diss}\geq0 ,
\end{equation}
where $\mathcal Q_{\rm diss}$ is the corresponding coarse-grained transfer
rate.

Using the expressions for $\rho_\theta$ and $p_\theta$, the left-hand side can
be written as
\begin{equation}
\dot\rho_\theta
+
3\mathcal H(\rho_\theta+p_\theta)
=
\frac{f^2}{a^2}\dot\theta
\left[
\ddot\theta
+
2\mathcal H\dot\theta
+
a^2\frac{\Lambda^4}{f^2}\sin\theta
\right].
\end{equation}
The energy transfer may therefore be represented at the level of
the local equation of motion by
\begin{equation}
\ddot\theta
+
2\mathcal H\dot\theta
+
a^2\frac{\Lambda^4}{f^2}\sin\theta
=
-R_{\rm diss},
\end{equation}
with
\begin{equation}
\mathcal Q_{\rm diss}
=
\frac{f^2}{a^2}\dot\theta\,R_{\rm diss}.
\end{equation}
Since $\mathcal Q_{\rm diss}\geq0$, the dissipative backreaction must satisfy
\begin{equation}
\label{FRLW_energy_loss_cond}
\dot\theta\,R_{\rm diss}\geq0 .
\end{equation}
Dissipative energy loss should occur for either sign of $\dot\theta$, so
Eq.~(\ref{FRLW_energy_loss_cond}) requires $R_{\rm diss}$ to change sign with
$\dot\theta$. In the
adiabatic regime, higher time-derivative terms are suppressed, and near
$\dot\theta=0$ the analytic form is therefore
\begin{equation}
R_{\rm diss}
=
\Gamma_{\rm eff}\dot\theta
+
O(\dot\theta^3),
\qquad
\Gamma_{\rm eff}\geq0 .
\end{equation}
Keeping only this leading adiabatic contribution leads to
\begin{equation}
\ddot\theta
+
\left(2\mathcal H+\Gamma_{\rm eff}\right)\dot\theta
+
a^2\frac{\Lambda^4}{f^2}\sin\theta
=
0 ,
\end{equation}
which is the FLRW local effective equation used in the main text,
Eq.~(\ref{Eq_conf}).

\printbibliography

@book{Lyth:2009zz,
    author = "Lyth, David H. and Liddle, Andrew R.",
    title = "{The primordial density perturbation: Cosmology, inflation and the origin of structure}",
    year = "2009"
}

@article{Greiner:1996dx,
    author = "Greiner, Carsten and Muller, Berndt",
    title = "{Classical fields near thermal equilibrium}",
    eprint = "hep-th/9605048",
    archivePrefix = "arXiv",
    reportNumber = "DUKE-TH-96-99",
    doi = "10.1103/PhysRevD.55.1026",
    journal = "Phys. Rev. D",
    volume = "55",
    pages = "1026--1046",
    year = "1997"
}

@article{Berera:2001gs,
    author = "Berera, Arjun and Ramos, Rudnei O.",
    title = "{The Affinity for scalar fields to dissipate}",
    eprint = "hep-ph/0101049",
    archivePrefix = "arXiv",
    doi = "10.1103/PhysRevD.63.103509",
    journal = "Phys. Rev. D",
    volume = "63",
    pages = "103509",
    year = "2001"
}

@article{Moss:2006gt,
    author = "Moss, Ian G and Xiong, Chun",
    title = "{Dissipation coefficients for supersymmetric inflatonary models}",
    eprint = "hep-ph/0603266",
    archivePrefix = "arXiv",
    month = "3",
    year = "2006"
}

@article{Berera:2008ar,
    author = "Berera, Arjun and Moss, Ian G. and Ramos, Rudnei O.",
    title = "{Warm Inflation and its Microphysical Basis}",
    eprint = "0808.1855",
    archivePrefix = "arXiv",
    primaryClass = "hep-ph",
    doi = "10.1088/0034-4885/72/2/026901",
    journal = "Rept. Prog. Phys.",
    volume = "72",
    pages = "026901",
    year = "2009"
}

@article{Buldgen:2019dus,
    author = "Buldgen, Gilles and Drewes, Marco and Kang, Jin U. and Mun, Ui Ri",
    title = "{General Markovian equation for scalar fields in a slowly evolving background}",
    eprint = "1912.02772",
    archivePrefix = "arXiv",
    primaryClass = "hep-ph",
    reportNumber = "CP3-19-54",
    doi = "10.1088/1475-7516/2022/05/039",
    journal = "JCAP",
    volume = "05",
    number = "05",
    pages = "039",
    year = "2022"
}

@article{Cline-published,
    author = "Cline, James M.",
    title = "{Comment on {\textquotedblleft}Spontaneous baryosynthesis with large initial phase{\textquotedblright}}",
    eprint = "2512.18516",
    archivePrefix = "arXiv",
    primaryClass = "hep-ph",
    doi = "10.1016/j.nuclphysb.2026.117524",
    journal = "Nucl. Phys. B",
    volume = "1029",
    pages = "117524",
    year = "2026"
}

@article{Belotsky:2018wph,
    author = "Belotsky, Konstantin M. and Dokuchaev, Vyacheslav I. and Eroshenko, Yury N. and Esipova, Ekaterina A. and Khlopov, Maxim Yu. and Khromykh, Leonid A. and Kirillov, Alexander A. and Nikulin, Valeriy V. and Rubin, Sergey G. and Svadkovsky, Igor V.",
    title = "{Clusters of primordial black holes}",
    eprint = "1807.06590",
    archivePrefix = "arXiv",
    primaryClass = "astro-ph.CO",
    doi = "10.1140/epjc/s10052-019-6741-4",
    journal = "Eur. Phys. J. C",
    volume = "79",
    number = "3",
    pages = "246",
    year = "2019"
}

@article{Barenboim_2019,
    author = "Barenboim, Gabriela and Park, Wan-Il",
    title = "{Spontaneous baryogenesis in spiral inflation}",
    eprint = "1901.05799",
    archivePrefix = "arXiv",
    primaryClass = "hep-ph",
    reportNumber = "IFIC/19-11",
    doi = "10.1140/epjc/s10052-019-6970-6",
    journal = "Eur. Phys. J. C",
    volume = "79",
    number = "6",
    pages = "456",
    year = "2019"
}

@article{Buchmuller:2000as,
    author = "Buchmuller, W. and Plumacher, M.",
    title = "{Neutrino masses and the baryon asymmetry}",
    eprint = "hep-ph/0007176",
    archivePrefix = "arXiv",
    reportNumber = "DESY-99-187, UPR-892-T",
    doi = "10.1142/S0217751X00002937",
    journal = "Int. J. Mod. Phys. A",
    volume = "15",
    pages = "5047--5086",
    year = "2000"
}

@article{Harvey:1990qw,
    author = "Harvey, Jeffrey A. and Turner, Michael S.",
    title = "{Cosmological Baryon and Lepton Number in the Presence of Electroweak Fermion Number Violation}",
    reportNumber = "FERMILAB-PUB-90-049-A, EFI-90-33",
    doi = "10.1103/PhysRevD.42.3344",
    journal = "Phys. Rev. D",
    volume = "42",
    pages = "3344--3349",
    year = "1990"
}

@article{Arbuzova_2016,
    author = "Arbuzova, E. V. and Dolgov, A. D. and Novikov, V. A.",
    title = "{General properties and kinetics of spontaneous baryogenesis}",
    eprint = "1607.01247",
    archivePrefix = "arXiv",
    primaryClass = "astro-ph.CO",
    doi = "10.1103/PhysRevD.94.123501",
    journal = "Phys. Rev. D",
    volume = "94",
    number = "12",
    pages = "123501",
    year = "2016"
}

@article{PhysRevLett.93.201301,
    author = "Davoudiasl, Hooman and Kitano, Ryuichiro and Kribs, Graham D. and Murayama, Hitoshi and Steinhardt, Paul J.",
    title = "{Gravitational baryogenesis}",
    eprint = "hep-ph/0403019",
    archivePrefix = "arXiv",
    doi = "10.1103/PhysRevLett.93.201301",
    journal = "Phys. Rev. Lett.",
    volume = "93",
    pages = "201301",
    year = "2004"
}

@inproceedings{Arbuzova:2017zby,
    author = "Arbuzova, E. V. and Dolgov, A. D.",
    title = "{Problems of spontaneous and gravitational baryogenesis}",
    booktitle = "{18th Lomonosov Conference on Elementary Particle Physics}",
    eprint = "1712.04627",
    archivePrefix = "arXiv",
    primaryClass = "hep-ph",
    doi = "10.1142/9789811202339_0059",
    pages = "309--313",
    year = "2019"
}

@article{Vennin_2015,
    author = "Vennin, Vincent and Starobinsky, Alexei A.",
    title = "{Correlation Functions in Stochastic Inflation}",
    eprint = "1506.04732",
    archivePrefix = "arXiv",
    primaryClass = "hep-th",
    doi = "10.1140/epjc/s10052-015-3643-y",
    journal = "Eur. Phys. J. C",
    volume = "75",
    pages = "413",
    year = "2015"
}

@article{Starobinsky:1986fx,
    author = "Starobinsky, Alexei A.",
    title = "{STOCHASTIC DE SITTER (INFLATIONARY) STAGE IN THE EARLY UNIVERSE}",
    doi = "10.1007/3-540-16452-9_6",
    journal = "Lect. Notes Phys.",
    volume = "246",
    pages = "107--126",
    year = "1986"
}

@article{KUZMIN,
    author = "Kuzmin, V. A.",
    title = "{CP-noninvariance and baryon asymmetry of the universe}",
    journal = "Pisma Zh. Eksp. Teor. Fiz.",
    volume = "12",
    number = "6",
    pages = "335--337",
    year = "1970"
}

@article{Sakharov:1967dj,
    author = "Sakharov, A. D.",
    title = "{Violation of CP Invariance, C asymmetry, and baryon asymmetry of the universe}",
    doi = "10.1070/PU1991v034n05ABEH002497",
    journal = "Pisma Zh. Eksp. Teor. Fiz.",
    volume = "5",
    pages = "32--35",
    year = "1967"
}

@article{Sakharov_2021,
    author = "Sakharov, Alexander S. and Eroshenko, Yury N. and Rubin, Sergey G.",
    title = "{Looking at the NANOGrav signal through the anthropic window of axionlike particles}",
    eprint = "2104.08750",
    archivePrefix = "arXiv",
    primaryClass = "hep-ph",
    doi = "10.1103/PhysRevD.104.043005",
    journal = "Phys. Rev. D",
    volume = "104",
    number = "4",
    pages = "043005",
    year = "2021"
}

@article{article,
author = {Guo, Shu-Yuan and Liu, Xuewen and Zhu, Bin and Khlopov, M. and Wu, Yongcheng},
year = {2024},
month = {11},
pages = {1-8},
title = {Footprints of axion-like particle in pulsar timing array data and James Webb Space Telescope observations},
volume = {67},
journal = {Science China Physics, Mechanics \& Astronomy},
doi = {10.1007/s11433-024-2445-1}
}

@article{Freese:1990rb,
    author = "Freese, Katherine and Frieman, Joshua A. and Olinto, Angela V.",
    title = "{Natural inflation with pseudo - Nambu-Goldstone bosons}",
    reportNumber = "FERMILAB-PUB-90-177-A",
    doi = "10.1103/PhysRevLett.65.3233",
    journal = "Phys. Rev. Lett.",
    volume = "65",
    pages = "3233--3236",
    year = "1990"
}

@article{dos_Santos_2024,
    author = "Dos Santos, F. B. M. and Rodrigues, G. and Rodrigues, J. G. and de Souza, R. and Alcaniz, J. S.",
    title = "{Is natural inflation in agreement with CMB data?}",
    eprint = "2312.12286",
    archivePrefix = "arXiv",
    primaryClass = "astro-ph.CO",
    doi = "10.1088/1475-7516/2024/03/038",
    journal = "JCAP",
    volume = "03",
    pages = "038",
    year = "2024"
}

@article{Planck:2018jri,
    author = "Akrami, Y. and others",
    collaboration = "Planck",
    title = "{Planck 2018 results. X. Constraints on inflation}",
    eprint = "1807.06211",
    archivePrefix = "arXiv",
    primaryClass = "astro-ph.CO",
    doi = "10.1051/0004-6361/201833887",
    journal = "Astron. Astrophys.",
    volume = "641",
    pages = "A10",
    year = "2020"
}

@article{Inomata:2018htm,
    author = "Inomata, Keisuke and Kawasaki, Masahiro and Kusenko, Alexander and Yang, Louis",
    title = "{Big Bang Nucleosynthesis Constraint on Baryonic Isocurvature Perturbations}",
    eprint = "1806.00123",
    archivePrefix = "arXiv",
    primaryClass = "astro-ph.CO",
    reportNumber = "IMPU-18-0101, IMPU 18-0101",
    doi = "10.1088/1475-7516/2018/12/003",
    journal = "JCAP",
    volume = "12",
    pages = "003",
    year = "2018"
}

@article{Montefalcone_2023,
    author = "Montefalcone, Gabriele and Aragam, Vikas and Visinelli, Luca and Freese, Katherine",
    title = "{Observational constraints on warm natural inflation}",
    eprint = "2212.04482",
    archivePrefix = "arXiv",
    primaryClass = "gr-qc",
    doi = "10.1088/1475-7516/2023/03/002",
    journal = "JCAP",
    volume = "03",
    pages = "002",
    year = "2023"
}

@article{Alam_2024,
    author = "Alam, Khursid and Dutta, Koushik and Jaman, Nur",
    title = "{CMB constraints on natural inflation with gauge field production}",
    eprint = "2405.10155",
    archivePrefix = "arXiv",
    primaryClass = "astro-ph.CO",
    doi = "10.1088/1475-7516/2024/12/015",
    journal = "JCAP",
    volume = "12",
    pages = "015",
    year = "2024"
}

@article{Planck:2018vyg,
    author = "Aghanim, N. and others",
    collaboration = "Planck",
    title = "{Planck 2018 results. VI. Cosmological parameters}",
    eprint = "1807.06209",
    archivePrefix = "arXiv",
    primaryClass = "astro-ph.CO",
    doi = "10.1051/0004-6361/201833910",
    journal = "Astron. Astrophys.",
    volume = "641",
    pages = "A6",
    year = "2020",
    note = "[Erratum: Astron.Astrophys. 652, C4 (2021)]"
}

@article{Cohen:1988kt,
    author = "Cohen, Andrew G. and Kaplan, David B.",
    title = "{Spontaneous Baryogenesis}",
    reportNumber = "HUTP-88/A016",
    doi = "10.1016/0550-3213(88)90134-4",
    journal = "Nucl. Phys. B",
    volume = "308",
    pages = "913--928",
    year = "1988"
}

@article{Cohen:1987vi,
    author = "Cohen, Andrew G. and Kaplan, David B.",
    title = "{Thermodynamic Generation of the Baryon Asymmetry}",
    reportNumber = "HUTP-87/A061",
    doi = "10.1016/0370-2693(87)91369-4",
    journal = "Phys. Lett. B",
    volume = "199",
    pages = "251--258",
    year = "1987"
}

@article{Dolgov:1991fr,
    author = "Dolgov, A. D.",
    title = "{NonGUT baryogenesis}",
    reportNumber = "YITP-K-940",
    doi = "10.1016/0370-1573(92)90107-B",
    journal = "Phys. Rept.",
    volume = "222",
    pages = "309--386",
    year = "1992"
}

@book{Khlopov:2004rw,
    author = "Khlopov, M. Yu. and Rubin, S. G.",
    title = "{Cosmological pattern of microphysics in the inflationary universe}",
    doi = "10.1007/978-1-4020-2650-8",
    publisher = "Springer Dordrecht",
    year = "2004"
}

@article{Barrow:2022gsu,
    author = "Barrow, J. L. and others",
    title = "{Theories and Experiments for Testable Baryogenesis Mechanisms: A Snowmass White Paper}",
    eprint = "2203.07059",
    archivePrefix = "arXiv",
    primaryClass = "hep-ph",
    month = "3",
    year = "2022"
}

@article{Dolgov_1997,
    author = "Dolgov, Alexandre and Freese, Katherine and Rangarajan, Raghavan and Srednicki, Mark",
    title = "{Baryogenesis during reheating in natural inflation and comments on spontaneous baryogenesis}",
    eprint = "hep-ph/9610405",
    archivePrefix = "arXiv",
    reportNumber = "ACT-09-96, TAC-96-015, UCSBTH-96-19, UM-AC-96-05",
    doi = "10.1103/PhysRevD.56.6155",
    journal = "Phys. Rev. D",
    volume = "56",
    pages = "6155--6165",
    year = "1997"
}

@article{Dolgov_1995,
    author = "Dolgov, Alexandre and Freese, Katherine",
    title = "{Calculation of particle production by Nambu Goldstone bosons with application to inflation reheating and baryogenesis}",
    eprint = "hep-ph/9410346",
    archivePrefix = "arXiv",
    reportNumber = "UM-AC-94-35",
    doi = "10.1103/PhysRevD.51.2693",
    journal = "Phys. Rev. D",
    volume = "51",
    pages = "2693--2702",
    year = "1995"
}

@article{Simone_2017,
    author = "De Simone, Andrea and Kobayashi, Takeshi",
    title = "{Spontaneous baryogenesis without baryon isocurvature}",
    eprint = "1610.05783",
    archivePrefix = "arXiv",
    primaryClass = "hep-ph",
    reportNumber = "SISSA-51-2016-FISI",
    doi = "10.1088/1475-7516/2017/02/036",
    journal = "JCAP",
    volume = "02",
    pages = "036",
    year = "2017"
}

@article{Simone_2016,
    author = "De Simone, Andrea and Kobayashi, Takeshi",
    title = "{Cosmological Aspects of Spontaneous Baryogenesis}",
    eprint = "1605.00670",
    archivePrefix = "arXiv",
    primaryClass = "hep-ph",
    reportNumber = "SISSA-24-2016-FISI",
    doi = "10.1088/1475-7516/2016/08/052",
    journal = "JCAP",
    volume = "08",
    pages = "052",
    year = "2016"
}

@article{LINDE1982335,
    author = "Linde, Andrei D.",
    title = "{Scalar Field Fluctuations in Expanding Universe and the New Inflationary Universe Scenario}",
    reportNumber = "Print-82-0555 (LEBEDEV INST)",
    doi = "10.1016/0370-2693(82)90293-3",
    journal = "Phys. Lett. B",
    volume = "116",
    pages = "335--339",
    year = "1982"
}

\end{document}